\documentclass[letterpaper,english,reprint, aps, prb, groupedaddress, notitlepage]{revtex4-1}
\usepackage[T1]{fontenc}
\usepackage[latin9]{inputenc}
\usepackage{babel}
\usepackage{mathtools}
\usepackage{amsmath}
\usepackage{amssymb}
\usepackage{graphicx}
\usepackage[unicode=true]
 {hyperref}

\makeatletter

\pdfpageheight\paperheight
\pdfpagewidth\paperwidth

\providecommand{\tabularnewline}{\\}

\usepackage{pdfsync} 

\usepackage[table]{xcolor} 

\usepackage[transparent]{nicematrix} 

\newcommand{\znm}{ 
	\NiceMatrixOptions{
		first-row, first-col,
		code-for-first-col = \color{gray},
		code-for-first-row = \color{gray}}
}

\makeatother

\begin{document}
\global\long\def\ket#1{\left|#1\right\rangle }%

\global\long\def\bra#1{\left\langle #1\right|}%

\global\long\def\braket#1#2{\left\langle #1\middle|#2\right\rangle }%

\global\long\def\ketbra#1#2{\left|#1\vphantom{#2}\right\rangle \left\langle \vphantom{#1}#2\right|}%

\global\long\def\braOket#1#2#3{\left\langle #1\middle|#2\middle|#3\right\rangle }%

\global\long\def\dag#1{#1^{\dagger}}%

\global\long\def\kb#1#2{\left|#1\vphantom{#2}\right\rangle \left\langle \vphantom{#1}#2\right|}%

\global\long\def\kk{\rangle\!\rangle}%
\global\long\def\bb{\langle\!\langle}%

\global\long\def\normord#1{{:\mathrel{\mspace{1mu}#1\mspace{1mu}}:}}%

\global\long\def\isdef{\coloneqq}%

\global\long\def\m#1{\mathbf{#1}}%

\global\long\def\Tr{\operatorname{Tr}}%

\global\long\def\Span{\operatorname{span}}%

\global\long\def\argmin{\operatornamewithlimits{argmin}}%

\global\long\def\image{\operatorname{im}}%

\global\long\def\rank{\operatorname{rank}}%

\global\long\def\block{\operatorname{block}}%

\global\long\def\ddt{\frac{\mathrm{d}}{\mathrm{d}t}}%

\global\long\def\p{\partial}%

\global\long\def\d{\mathrm{d}}%

\global\long\def\dt{\mathrm{d}t}%

\title{Circuit quantum electrodynamics (cQED) with modular quasi-lumped models}
\author{Zlatko~K.~Minev}
\email{zlatko.minev@ibm.com; www.zlatko-minev.com}

\author{Thomas~G.~McConkey}
\author{Maika~Takita}
\author{Antonio~D.~Corcoles}
\author{Jay~M.~Gambetta}
\affiliation{IBM Quantum, IBM T.J. Watson Research Center, Yorktown Heights, US}
\date{\today}
\begin{abstract}
Extracting the Hamiltonian of interacting quantum-information processing
systems is a keystone problem in the realization of complex phenomena
and large-scale quantum computers. The remarkable growth of the field
increasingly requires precise, widely-applicable, and modular methods
that can model the quantum electrodynamics of the physical circuits,
including their more-subtle renormalization effects. Here, we present
a computationally-efficient method satisfying these criteria. The
method partitions a quantum device into compact lumped or quasi-distributed
cells. Each is first simulated individually. The composite system
is then reduced and mapped to a set of simple subsystem building blocks
and their pairwise interactions. The method operates within the quasi-lumped
approximation and, with no further approximation, systematically accounts
for constraints, couplings, parameter renormalizations, and non-perturbative
loading effects. We experimentally validate the method on large-scale,
state-of-the-art superconducting quantum processors. We find that
the full method improves the experimental agreement by a factor of
two over taking standard coupling approximations when tested on the
most sensitive and dressed Hamiltonian parameters of the measured
devices. 
\end{abstract}
\maketitle
Quantum phenomena offer a distinct advantage for information processing,
assuming we can faithfully design and realize the physical systems
underlying them. A leading platform to accomplish this goal has emerged
in the form of superconducting quantum technology \citep{Devoret2013,Gambetta2017,Krantz2019,Kjaergaard2020,Blais2020},
which employs macroscopic, lithographically-defined, and configurable
devices. Their design versatility, however, comes with
inherent challenges\textemdash parameter variability, a complicated design space,
and the difficulty of engineering their non-linear interactions. These
challenges pose a concern of central importance for the growth of
the field and have received a strong and growing interest in the form
of new design and quantization methods \citep{Nigg2012,Bourassa2012,Solgun2014,Solgun2015,Smith2016,Malekakhlagh2016-A2,Gely2017,Malekakhlagh2017-Cutoff-Free,Pechal2017,pyEPR,Parra-Rodriguez2018,Parra-Rodriguez2018a,Ansari2018,Krupko2018,Malekakhlagh2018,Solgun2017, Petrescu2019,You2019-Koch,DiPaolo2019,Gely2020,Ding2020,Kerman2020,Minev2020,Kyaw2020,Malekakhlagh2020,Menke2021}.
The rapidly growing pace of diversity, complexity, and scale of quantum
hardware \citep{Wallraff2004,Paik2011,Barends2013,Minev2013,FYan2016,Brecht2016,Rosenberg2017,Versluis2017,Naik2017,CharlesJamesNeill2017,Yan2018,Minev2019Nature,Corcoles2021}
urges for methods that are increasingly modular, widely-applicable,
yet ever-more precise. As such, these methods must closely incorporate details
of the layout, materials, and electromagnetic environment of the physical device, while keeping approximations to a minimum without sacrificing computational efficiency.

Here, we develop such a precise, modular method that operates at the physical-device level and that is suitable for a wide array of quantum devices.
The method builds on the quantization of lumped models \citep{Yurke1984,Devoret1995,Burkard2004},
which is more computationally efficient than full-wave methods \citep{Nigg2012,Solgun2015,Minev2020}.
The method also handles distributed resonant structures, such as co-planar
waveguide (CPW) resonators. Modularity is achieved in two ways. First,
the physical layout of the quantum device is systematically partitioned
into disjoint cells\textemdash physical blocks of the device. Each
cell can be independently simulated to extract its electromagnetic
parameters. Second, the effective circuit of the device is partitioned
into non-dynamical coupler elements and nodes and into subsystem
building blocks. A subsystem is selected based on two requirements:
i) it should constitute a well-understood, basic system in isolation
and ii) it should have sufficient parameter flexibility. The latter enables the faithful reduction and mapping of the larger device into
subsystems, without loss of information. The larger device model is
stitched together from the results of the cell simulations. It is
then reduced according to the subsystem partitions. In the process,
constraints and non-dynamical degrees of freedom associated with the
coupling structures are systematically eliminated; dressing of the
systems and their interactions because of this are accounted for.
The reduction is precise in that no approximations are made. All dressing and effective parameter renormalizations of the subsystems are tracked, which are mediated by the coupling structures.

We experimentally tested the method on two large-scale, superconducting
quantum processors \citep{Corcoles2021}, which employed transmon
qubits \citep{Koch2007} and CPW structures. Each qubit was connected
to four or five neighboring structures, spanning a wide range of coupling strengths. We observed renormalizations on the order of~25\%
for subsystem Hamiltonian parameters due to coupling dressing and
a series of smaller renormalizations due to unwanted, indirect couplings
and higher-order effects. 
We provide a detailed budget to account for the extent to which each model parameter influenced the results.
We compare the full method presented here to one that resorts to standard approximations, weak coupling and no dressing of spatial eigenfields. The full method presented here yields a factor of two improvement on the experimental agreement.
The method is applicable to the analysis of a broad class of quantum processors.
We have automated
it in the open-source project~\textsc{Qiskit Metal | for quantum
device design} \footnote{For an early version of the open-source code \citep{Qiskit_Metal} developed by the authors
of this manuscript, see \href{http://www.qiskit.org/metal}{qiskit.org/metal}.
The Qiskit Metal project builds on \href{http://http:/github.com/zlatko-minev/pyEPR}{github.com/zlatko-minev/pyEPR}.}.

\section{Partitioning a quantum device into interconnected component cells}

\label{sec:cell-model}

\begin{figure}
\begin{centering}
\includegraphics{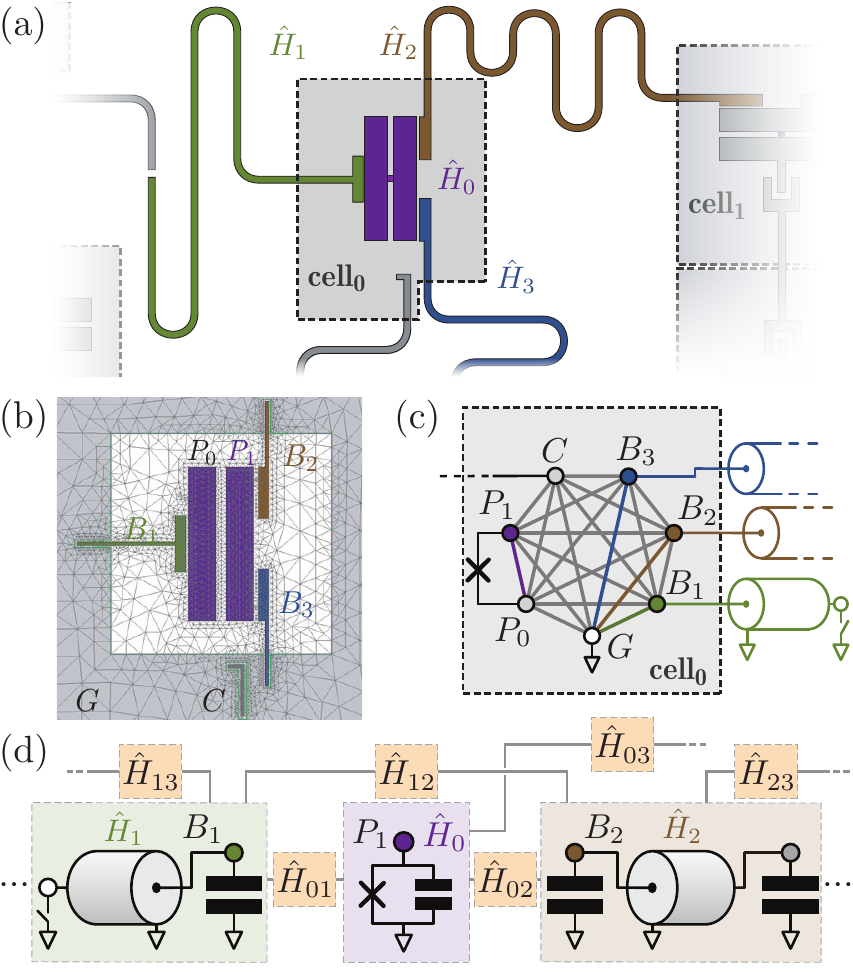}
\par\end{centering}
\caption{\label{fig:model}Method overview. (a) Illustration of an example quantum
processor layout (not-to-scale; partial). Center: transmon qubit subsystem (purple),
described by a dressed Hamiltonian~$\hat{H}_{0}$, connected to three co-planar waveguide (CPW) subsystems---a readout (green, $\hat H_1$) and two bus resonators (brown, $\hat H_2$, and blue, $\hat H_3$)---and a charge line (bottom, gray).
The device layout is partitioned into subsystems and cells.
(b) Example simulation model of~$\mathrm{cell}_0$, overlaid with its simulation mesh. 
The cell incorporates elements of multiple subsystems: qubit pads~$P_{0}$ and~$P_{1}$, CPW coupler pads and segments~$B_{1},B_{2},B_{3}$, and a segment of a CPW charge line~$C$.
(c) Partial schematic of the composite-system network depicting the nodes and elements of~$\mathrm{cell}_0$ and their connections to neighboring cells, such as those of the CPWs. 
The nodes of the cell are capacitively coupled by a fully-connected graph (thick line). 
A Josephson tunnel junction connects~$P_{0}$
to~$P_{1}$. Coupler nodes~$C$ and~$P_{0}$ (not colored) are eliminated.
(d) Depiction of the reduced, dressed subsystems---acting as building blocks---and their dressed interactions, described by the Hamiltonians~$\hat{H}_{01},\hat{H}_{02},\ldots$
}
\end{figure}

The physical layout of a quantum information processor, such as the
one depicted in Fig.~\ref{fig:model}(a), can be thought of as the
interconnected collection of quasi-independent subsystems and inter-system
couplers. Each subsystem is identified with a domain of the physical
layout and supports a set of quantized, potentially-anharmonic modes.
When considered in isolation, a subsystem could be one that is well
understood and whose parameters could be readily obtainable, using
analytical or numerical techniques. However, due to its embedding
in the larger device network and loading by system-system coupling
structures, its properties can be dressed and can significantly depend
on parts of the larger network. Generally, this dressing of the Hamiltonian
parameters is necessarily unavoidable and non-local. Even for relatively
moderate non-linear coupling strengths, the renormalizations of the
effective subsystem Hamiltonian and its spectrum can be significant;
for an experimental example of a large,~1~GHz mode dressing, see
Sec.~\ref{sec:Experiment}. Similarly, the interaction between two
subsystems is dressed in a potentially non-local manner
that can depend on more than the immediate coupling structure and
can be influenced by the physical layout more broadly. Faithfully
and systematically accounting for such effects in the increasingly
complex physical device layouts is important for improved experimental
agreement, see Fig.~\ref{fig:expeirment}.

To capture these effects, we aim to systematically and modularly study
the physical device layout by partitioning it into disjoint cells,
see Fig.~\ref{fig:model}(b). Each cell is first analyzed quasi-analytically
or using numerical quasi-static electromagnetic methods to extract
device parameters used to construct the full system Hamiltonian. Importantly,
cells do not correspond to subsystems, see Sec.~\ref{sec:Composite-model}.
A semi-classical model of the connected system is constructed and
reduced to a simpler, dressed model of the device, see Fig.~\ref{fig:model}(b)
and \ref{fig:model}(c). 

For a subsystem coupled to~$K$ neighbors, we explicitly construct
the composite-system Hamiltonian~$\hat{H}_{\mathrm{full}}$, within
the quasi-lumped approximation but with no further approximations,
in the pair-wise interaction form,

\begin{equation}
\hat{H}_{\mathrm{full}}=\hat{H}_{0}+\sum_{n=1}^{K}\hat{H}_{n}+\sum_{n=0}^{K-1}\sum_{m=n+1}^{K}\hat{H}_{nm}\;,\label{eq:H-full}
\end{equation}
where~$\hat{H}_{0}$ and~$\hat{H}_{n}$ are the dressed Hamiltonians
of the subsystem and its~$n$-th neighbor, respectively, and~$\hat{H}_{nm}$
is the dressed, bi-linear interaction Hamiltonian between the~$n$-th
and $m$-th systems. 

In the construction, the~$n$-th and~$m$-th system Hilbert spaces
are disjoint; hence,~$\left[\hat{H}_{n},\hat{H}_{m}\right]=0$. Nonetheless,
their dressed Hamiltonians~$\hat{H}_{n}$ and~$\hat{H}_{m}$ are
interdependent in terms of their parameters and physical layouts\textemdash due
to the coupling structures and dressing by the larger network. 

The dressed interaction Hamiltonian of the~$n$-th and~$m$-th systems
is
\begin{equation}
\hat{H}_{nm}=\hat{Q}_{n}\hat{Q}_{m}/\text{\ensuremath{C_{nm}^{\mathrm{eff}}}}+\hat{\Phi}_{n}\hat{\Phi}_{m}/\text{\ensuremath{L_{nm}^{\mathrm{eff}}}}\;,\label{eq:H-int}
\end{equation}
where the effective coupling capacitance and inductance are~$\text{\ensuremath{C_{nm}^{\mathrm{eff}}}}$
and~$L_{nm}^{\mathrm{eff}}$, respectively, and the generalized charge
and magnetic flux operators of the~$n$-th and~$m$-th systems involved
in the coupling are $\hat{Q}_{n}$, $\hat{Q}_{m}$, $\hat{\Phi}_{n}$,
and~$\hat{\Phi}_{m}$. For systems coupled by multiple physical degrees
of freedom, see the more general form of Eq.~\eqref{eq:H-int} obtained
in Eqs.~\eqref{eq:H-nm-explicit} and~\eqref{eq:Ceff-Leff-block}.

In the case of purely capacitive inter-system coupling (i.e., $1/L_{nm}^{\mathrm{eff}}=0$),
we conveniently reexpress the interaction in terms of the dimensionless
operators~$\hat{A}_{n}\isdef\hat{Q}_{n}/A_{n}$ and~$\hat{B}_{m}\isdef\hat{Q}_{m}/B_{m}$,
scaled by the choice scaling factors~$A_{n}$ and~$B_{m}$,
\begin{equation}
\hat{H}_{nm}=\hat{Q}_{n}\hat{Q}_{m}/\text{\ensuremath{C_{nm}^{\mathrm{eff}}}}=\hbar g_{nm}\hat{A}_{n}\hat{B}_{m}\;,\label{eq:Hnm-dimensionless}
\end{equation}
where the linear coupling energy~$\hbar g_{nm}\isdef A_{n}B_{m}/\text{\ensuremath{C_{nm}^{\mathrm{eff}}}}$.
When operating with continuous quadrature variables, such as~$\hat{A}_{n}=i(\hat{a}_{n}^{\dagger}-\hat{a}_{n})$,
and employing a harmonic basis, one typically employs the zero-point
quantum fluctuations of the quadrature with respect to the linearized
system; i.e.,~$\text{\ensuremath{A_{n}}}^{2}=\left(A_{n}^{\mathrm{ZPF}}\right){}^{2}\isdef\left\langle \hat{A}_{n}^{2}\right\rangle -\left\langle \hat{A}_{n}\right\rangle ^{2}$.
Sometimes in the case of a transmon qubit \citep{Koch2007}, especially
when interested in charge effects, one operates with wrapped or discrete-variable
quadratures. In the case of the discrete-variable Cooper pair number
operator~$\hat{A}_{n}=\hat{n}$, the scaling is $A_{n}=2e$, where~$e$
is the elementary electron charge.

Before proceeding to the general treatment and explicit construction
of Eqs.~\eqref{eq:H-full}\textendash \eqref{eq:CL-composition},
for definitiveness of example, consider the device and associated
reduction steps depicted in Fig.~\ref{fig:model}. The dressed Hamiltonian
of the~$k=1$ system, a co-planar-waveguide (CPW) resonator, is~$\hat{H}_{1}=\sum_{m=1}^{\infty}\hbar\omega_{1m}\left(\text{\ensuremath{\hat{a}_{1m}^{\dagger}\hat{a}_{1m}}}+\frac{1}{2}\right)$,
see Sec.~\ref{sec:LTL}. Due to the coupling and network dressings,
the normal mode frequencies~$\omega_{1m}$, annihilation operators~$\hat{a}_{1m}$,
and associated zero-point-quantum fluctuations of the physical degrees
of freedom are not dependent solely on the geometry and materials of the CPW
structure. They additionally include contributions arising
from the coupler structure and transmon qubit self-capacitance. Similarly,
the $k=0$ qubit system frequency~$\omega_{q}$ of~$\hat{H}_{0}$
is weakly interdependent with~$\omega_{1m}$ due to the semi-classical
hybridization; i.e., when one tunes~$\omega_{q}$, one has to be
careful to also retune~$\omega_{1m}$ for maximum accuracy. The coupling
parameters~$\text{\ensuremath{C_{nm}^{\mathrm{eff}}}}$ and~$L_{nm}^{\mathrm{eff}}$
are dressed not only by the~$n$-th and~$m$-th systems, but also
by couplers coupling the~$n$-th system to its other neighbors; e.g.,
the qubit-bus coupling can weakly influence the qubit-readout coupling.

In the following section, we capture these potentially non-perturbative
effects, see Sec.~\ref{sec:LTL}. Such a detailed construction is
required to treat larger and higher-order coupling effects and to
obtain improved experimental agreement, see Sec.~\ref{sec:Experiment}.
Additionally, effects such as impurity scattering on the distributed
CPW structures are accounted for\textemdash this is the cavity-QED
equivalent of a gauge-dependent diamagnetic~$A^{2}$ contribution
\citep{Malekakhlagh2016-A2}.

\section{Theory of the composite model}

\label{sec:Composite-model}

In this section, we present the theory of the general construction
of Eqs.~\eqref{eq:H-full} and~\eqref{eq:H-int} starting from the
analysis of disjoint partitions of the physical device layout. The
model is constructed to allow the mapping of the larger interconnected
network to a set of small, independent, subsystem building blocks\textemdash each
potentially well understood in isolation. Each such cell partition
can be independently simulated, thus decoupling the complexity of the
network and providing modular simulations for improved computational
efficiency. From the results of these independent simulations, we
construct the composite-system Lagrangian and then the quantum Hamiltonian
of the systems and interactions. In the process, we eliminate holonomic
constraints and non-dynamical, coupling degrees of freedom, and account
for their dressing of the systems and couplings. The treatment in
this section, for generality, is more formal and abstract. However,
in practice, significant simplifications occur due to the typically
diagonal, block, or sparse structure of the circuit matrices. 

We simultaneously employ two different types of partitions of the
device. The first partitions the physical layout of the device into~$N_{\mathrm{cell}}+1$
strictly disjoint cell modules\textemdash each of which can be independently
simulated or analyzed to extract effective circuit parameters. The
second partition operates at the level of the effective quasi-lumped
schematic of the device. It partitions the device network nodes and
elements into~$K+2$ subsets of each. One of the~$K+2$ subsets
is for all non-system couplers. The remaining~$K+1$ are for the~$K+1$
subsystems. The cell and system partitions of the composite circuit
are distinct from each other. This is illustrated by the  qubit cell
of Fig.~\ref{fig:model}(b). It incorporates the transmon qubit elements
(metal pads~$P_{0}$ and~$P_{1}$ in a cutout of the ground plane~$G$)
but also elements from three neighboring non-qubit systems (structures~$B_{1},B_{2},B_{3,}$
and~$C$).

Let~$\mathcal{N}_{\mathrm{full}}$ and~$\mathcal{B}_{\mathrm{full}}$
denote the sets of all nodes and elements in the device network, respectively.
To partition into cells, the elements of~$\mathcal{B}_{\mathrm{full}}$
are disjointly distributed among~$N_{\mathrm{cell}}+1$ sets. The
nodes in~$\mathcal{N}_{\mathrm{full}}$ are however distributed among
sets with potential overlap. If we denote the set of all non-datum
nodes assigned to the~$n$-th cell as~$\mathcal{N}_{n}^{\mathrm{cell}}$,
then~$\mathcal{N}_{\mathrm{full}}=\mathcal{N}_{g}\cup\bigcup_{n=0}^{N_{\mathrm{cell}}}\mathcal{N}_{n}^{\mathrm{cell}}$,
where~$\mathcal{N}_{g}$ is the single-element set comprising just
the circuit datum (ground node). The intersection~$\mathcal{N}_{n}^{\mathrm{cell}}\cap\mathcal{N}_{m}^{\mathrm{cell}}$
can be non-empty. Each cell can incorporate the datum. To partition
into subsystems,~$\mathcal{B}_{\mathrm{full}}$ and~$\mathcal{N}_{\mathrm{full}}$
are broken up into strictly disjoint sub-sets. If~$\mathcal{N}_{k}$
denotes the set of all non-datum nodes assigned to the~$k$-th system
and~$\mathcal{N}_{\mathrm{couple}}$ denotes the set of all non-system,
non-datum nodes, then~$\mathcal{N}_{\mathrm{full}}=\mathcal{N}_{g}\cup\mathcal{N}_{\mathrm{couple}}\cup\bigcup_{k=0}^{K}\mathcal{N}_{k}$,
where~$\mathcal{N}_{k}\cap\mathcal{N}_{k'}=\mathcal{N}_{k}\cap\mathcal{N}_{\mathrm{couple}}=\emptyset$
for~$k\neq k'$. In the case of a continuous subsystem,~$\mathcal{N}_{k}$
comprises a continuum of nodes. 

The assignment of nodes and elements to a subsystem is determined
by the design intention to realize quasi-independent subsystems, incorporating
dynamical degrees of freedom that support quantized modes. A subsystem
can also incorporate non-dynamical nodes and zero-frequency modes;
e.g., a transmission line open at both ends. The system's non-dynamical
degrees will be preserved in the following reduction. This feature
provides helpful flexibility that allows the mapping of the larger
network to a set of known building blocks. 

A second, necessary condition for a subsystem is that it admits sufficient
parameter flexibility in its definition. It must be able to admit
the renormalization of the larger network. For example, the simple,
ideal transmission line is not a suitable subsystem candidate. It
cannot admit in its homogenous construction the inhomogeneity introduced
by the impurity scattering effect of a coupler structure. The line
must be allowed to incorporate the inhomogeneity in line parameters
due to the coupler structure in its definition; see Fig.~\ref{fig:ltl-model}
for an example.

Nodes and elements not part of any subsystem are assigned to the coupler
sets. The coupler nodes should all be non-dynamical\textemdash they
do not support quantized modes on their own. A sufficient condition
to identify a non-dynamical node is that it is touched by only inductive
or capacitive elements, but not by elements of both classes. 

Inductive elements can be linear or non-linear. A non-linear inductive
dipole is a two-terminal purely-inductive sub-circuit, such as a Josephson
tunnel junction, flux-biased SQUID \citep{Zimmerman1966,Clarke2004},
SNAIL \citep{Frattini2018}, or a more complicated composite sub-circuit.
We describe such dipoles using the formulation of Ref.~\onlinecite{Minev2020}.
The~$j$-th non-linear dipole is fully characterized by a known energy
function~$\mathcal{E}_{j}\text{\ensuremath{\left(\Phi_{j};\Phi_{j}^{\mathrm{ext}}\right)}}$
of the generalized magnetic flux~$\Phi_{j}$ across it and any external
bias parameters~$\Phi_{j}^{\mathrm{ext}}$, such as the magnetic flux
of a d.c. voltage bias. The dipole can also have an intrinsic capacitance~$C_{j}$
that spans its two terminals. The energy function of the dipole intrinsic
capacitance is~$\frac{1}{2}C_{j}\dot{\Phi}_{j}^{2}$.

With respect to the circuit operating point \citep{Minev2019-Thesis,Minev2020},
one can partition the inductive energy into strictly linear and non-linear
contributions
\begin{subequations}
\label{eq:ejlin-nl}
\begin{align}
\mathcal{E}_{j}^{\mathrm{lin}}\text{\ensuremath{\left(\Phi_{j};\Phi_{j}^{\mathrm{ext}}\right)}} & \isdef\frac{1}{2}L_{j}^{-1}\left(\Phi_{j}^{\mathrm{ext}}\right)\Phi_{j}^{2}\;,\\
\mathcal{E}_{j}^{\mathrm{nl}}\text{\ensuremath{\left(\Phi_{j};\Phi_{j}^{\mathrm{ext}}\right)}} & \isdef\mathcal{E}_{j}\left(\Phi_{j};\Phi_{j}^{\mathrm{ext}}\right)-\frac{1}{2}L_{j}^{-1}\left(\Phi_{j}^{\mathrm{ext}}\right)\Phi_{j}^{2}\;,
\end{align}
\end{subequations}
respectively, where~$L_{j}^{-1}\left(\Phi_{j}^{\mathrm{ext}}\right)$
is the linear-response inductance of the dipole at the bias point.
For example, for a SQUID dipole, $\mathcal{E}_{j}\left(\Phi_{j};\Phi_{j}^{\mathrm{ext}}\right)=-E_{j}\left(\Phi_{j}^{\mathrm{ext}}\right)\cos\left(\Phi_{j}/\phi_{0}\right)$,
where~$E_{j}$ is the effective Josephson energy as a function of
the bias and~$\phi_{0}\isdef\hbar/2e$. For a Josephson tunnel junction,~$L_{j}$
is simply the Josephson inductance, which can be computed from the
Ambegaokar-Baratoff expression and room-temperature resistance measurements
of the junction \citep{Gloos2000}. Henceforth, we make the bias argument~$\Phi_{j}^{\mathrm{ext}}$
implicit. 

The energy functions of the~$J$ non-linear dipoles and the structure
and topology of the~$N_{\mathrm{cell}}+1$ cells are known. The circuit
parameters of the linear elements contained in~$\mathcal{B}_{\mathrm{full}}$
can be extracted using analytical results \citep{Wolff2006} or numerical
techniques, see Sec.~\ref{sec:Lumped-cell-model}. Information about
the linear part of the circuit of the~$n$-th cell can be organized
in the cell's geometric capacitance and inverse inductance matrices~$\m C_{n,\mathrm{cell}}'$
and~${\m L'}_{n,\mathrm{cell}}^{-1}$, respectively. We define these
matrices with respect to the node-to-datum generalized magnetic fluxes
of the circuit \citep{Devoret1995,Minev2020}. In the case of a cell
that can be considered in the lumped regime, $\m C_{n,\mathrm{cell}}'$
is the reduced Maxwell capacitance matrix, see Eq.~\eqref{eq:Maxwell-mat-reduced}.
In this case,~$\m C_{n,\mathrm{cell}}'$ represents a fully connected
graph, see Fig.~\ref{fig:model}(c). 

We incorporate~$C_{j}$ and~$L_{j}$ of the non-linear dipoles into~$\m C_{n,\mathrm{cell}}'$
and~${\m L'}_{n,\mathrm{cell}}^{-1}$ to define the total cell linear
capacitance~$\m C_{n,\mathrm{cell}}$ and inverse inductance~$\m L{}_{n,\mathrm{cell}}^{-1}$
matrices. In terms of these, the capacitance and inverse inductance
matrices of the composite-system are
\begin{equation}
\m{C_{n}}=\sum_{n=0}^{N_{\mathrm{cell}}}\m C_{n,\mathrm{cell}}\quad\text{and}\quad\m L_{\m n}^{-1}=\sum_{n=0}^{N_{\mathrm{cell}}}\m L_{n,\mathrm{cell}}^{-1}\;,\label{eq:CL-composition}
\end{equation}
respectively. They are real, symmetric, and nearly block diagonal.
The matrix~$\m{C_{n}}$ is guaranteed to be positive semi-definite,
but due to the inclusion of~$L_{j}$, $\m L_{\m n}^{-1}$ does not
have this guarantee. Typically,~$\m{\m L_{n}}^{-1}$ is sparse and
rank deficient. In terms of system, not cell, partitions,~$\m{C_{n}}$
and~$\m L_{\m n}^{-1}$ have a simple block structure. All subsystems
are described by blocks in the diagonals of~$\m{C_{n}}$ and~$\m L_{\m n}^{-1}$.
The only off-diagonal blocks in the two matrices are due to coupling
elements spanning subsystems. Coupler nodes~$\mathcal{N}_{\mathrm{couple}}$
also introduce diagonal blocks in the two matrices. 

To describe the physical model, we introduce the column vector of
all node-to-datum fluxes~$\m{\Phi_{n}}$, each entry of which is
associated with one unique node of~$\mathcal{N}_{\mathrm{full}}-\mathcal{N}_{g}$.
The Lagrangian of the composite system is
\begin{equation}
\mathcal{L}_{\mathrm{n}}\left(\m{\Phi_{n}},\m{\dot{\Phi}_{n}}\right)=\frac{1}{2}\m{\dot{\Phi}_{n}}^{\intercal}\m{C_{n}}\m{\dot{\Phi}_{n}}-\frac{1}{2}\m{\Phi_{n}}^{\intercal}\m{L_{n}}^{-1}\m{\Phi_{n}}-\sum_{j=1}^{J}\mathcal{E}_{j}^{\mathrm{nl}}\left(\Phi_{j}\right)\;,
\end{equation}
where~$J$ denotes the total number of non-linear dipoles in the
network. The Lagrangian is potentially singular due to the rank deficiency
of~$\m L_{\m n}^{-1}$ and that of~$\m{C_{n}}$. The singularity
leads to dynamics on a constrained submanifold of phase space. In
quantization, it is this reduced, constrained phase space that provides
the required physical Poisson structure required for canonical quantization
\citep{Dirac1982-Book}. 

Before eliminating the coupler constraints on the phase space associated
with~$\mathcal{L}_{\mathrm{n}}$, we first rotate the Lagrangian
basis~$\m{\Phi_{n}}$ to place all non-linear dipole fluxes explicitly
in the basis. The flux of the~$j$-th dipole~$\Phi_{j}=\Phi_{n_{2}}-\Phi_{n_{1}}$,
where~$\Phi_{n_{2}}$ and~$\Phi_{n_{1}}$ are the node-to-datum
fluxes of its two nodes. We construct the simple, linear transformation~$\m{S_{n}}^{-1}$
with elements in~$\left\{ -1,0,1\right\} $ to rotate the basis according
to~$\m{\Phi}=\m{S_{n}}^{-1}\m{\m{\Phi_{n}}}$ such that~$\m{\Phi}$
explicitly contains all junction fluxes~$\Phi_{j}$.  In the basis~$\m{\Phi}$,
the transformed capacitance and inverse inductance matrices are~$\m C=\m{S_{n}}^{\intercal}\m{C_{n}}\m{S_{n}}$
and~$\m L^{-1}=\m{S_{n}}^{\intercal}\m{L_{n}}^{-1}\m{S_{n}}$, respectively. 

To eliminate constraints due to the singularity of~$\m L^{-1}$,
we select all~$r$ coupler fluxes or linear combinations of fluxes in the
kernel space of~$\m L^{-1}$. For example, these include the fluxes
associated with all nodes in~$\mathcal{N}_{\mathrm{couple}}$ that
have only capacitive elements touching them. We construct the~$N\times r$
matrix formed by joining the~$r$ flux vectors~$\m s_{1},\m s_{2},\ldots,\m s_{r}$
defined in the~$\m{\Phi}$ basis as~$\m S_{r}\isdef\left[\m s_{1},\ldots,\m s_{r}\right]$.
The flux-vector image space of~$\m S_{r}$ is a non-dynamical subspace
of the inductors; i.e.,~$\m L^{-1}\m S_{r}=\m 0_{Nr}$, where~$\m 0_{Nr}$
is the~$N\times r$ matrix of all zeros. The image space of~$\m S_{r}$
is not necessarily the full kernel space of~$\m L^{-1}$; i.e., $\Span\left(\m S_{r}\right)\subseteq\ker\left(\m L^{-1}\right)$.

We purposefully do not fully reduce the dynamics to only the image
space of~$\m L^{-1}$ at this stage. This allows the composite network
model to be reduced to known subsystems. These subsystems, treated
as building blocks of the device, can themselves incorporate a singular
Lagrangian. A simple example is that of the open-ended transmission
line, which effectively has one less inductor than capacitor and hence
a kernel space of~$\m L^{-1}$ with dimension one, which leads to
one zero-frequency solution. To map to this known problem, which itself
handles its own singularity, we retain the singularity in its definition
by not including the associated subspace in the span of~$\m S_{r}$.

We denote the complement of~$\m S_{r}$ as~$\m S_{k}$, an~$N\times\left(N-r\right)$
matrix whose image is the image space of~$\m L^{-1}$ together with
subspace of the kernel of~$\m L^{-1}$ that is not spanned by~$\m S_{r}$;
i.e.,~$\mathbb{R}^{N}=\Span\left(\m S_{r}\right)+\Span\left(\m S_{k}\right)$.
Its columns can be selected mostly if not entirely from those of
the identity matrix.

We partition the degrees of freedom~$\m{\Phi}$ of the composite
system into the~$r$ non-dynamical degrees of freedom~$\m{\Phi_{r}}\isdef\m S_{r}^{\intercal}\m{\Phi}$
and the remaining ones~$\m{\Phi_{k}}\isdef\m S_{k}^{\intercal}\m{\Phi}$.
For the non-dynamical coordinates, we impose the standard zero initial
condition~$\m{\Phi_{r}}\left(t_{0}\right)=\left(\ddt\m{\Phi_{r}}\right)\left(t_{0}\right)=\m 0_{r}$,
where~$t_{0}$ denotes the initial time of the circuit and~$\m 0_{r}$
is the length-$r$ column vector of all zeros. Using this, we solve
for the~$N$ degrees comprising~$\m{\Phi}$ in terms of the~$N-r$
reduced dynamical degrees~$\m{\Phi_{k}}$, and find~$\m{\Phi}=\m S_{k}\m{\Phi_{k}}$.
In the reduced basis, the reduced capacitance and inverse inductance
matrices are
\begin{subequations}
\label{eq:reduced-Li-C}
\begin{align}
\m{L_{k}}^{-1} & \isdef\m S_{k}^{\intercal}\m L^{-1}\m S_{k}\;,\\
\m{C_{k}} & \isdef\m S_{k}^{\intercal}\left(\m C-\m C\m S_{r}\left(\m S_{r}^{\intercal}\m C\m S_{r}\right)^{-1}\m S_{r}\m C\right)\m S_{k}\;,\label{eq:c-reduced1}
\end{align}
\end{subequations}
respectively. The matrix~$\m{C_{k}}$ can be seen as a Schur complement. 

In Eq.~\eqref{eq:reduced-Li-C}, we have eliminated the holonomic
constraints presented by inductive coupler nodes and contracted the
problem from~$N$ to~$N-r$ degrees of freedom~$\m{\Phi_{k}}$.
The reduction of the coupler nodes renormalizes system parameters
subject to the transformation~$\m S_{k}$ and Eq.~\eqref{eq:reduced-Li-C}.
In particular, the effective capacitance matrix can be strongly dressed,
due to the term containing~$\left(\m S_{r}^{\intercal}\m C\m S_{r}\right)^{-1}$.
This is only a first dressing. The capacitance matrix can be dressed
two more times in the following.

\paragraph{}

The second renormalization step parallels the first, but this time
with the roles of the capacitance and inductance matrices flipped.
Namely, if there are coupler flux combinations in the kernel space~$\ker\left(\m{C_{k}}\right)$,
we repeat the above basis partition protocol to eliminate these with
respect to~$\m{C_{k}}$ (rather than~$\m L^{-1}$, as above). An
example of a coupler flux in~$\ker\left(\m{C_{k}}\right)$ is the
flux of a coupler node that is only touched by inductors and no capacitors.
Since the steps repeat, for brevity, we do not explicitly detail the
process again. Instead, we proceed by assuming the step has been carried
out, all coupler nodes in~$\ker\left(\m{C_{k}}\right)$ and in~$\ker\left(\m L^{-1}\right)$
have been eliminated. To avoid new notation, we redefine~$\m{\Phi_{k}}$,
$\m{L_{k}}^{-1}$ and~$\m{C_{k}}$ to denote in the following the
doubly reduced flux basis vector and circuit matrices and~$N-r$
to denote the length of~$\m{\Phi_{k}}$.

The generalized, reduced charge vector is~$\m{Q_{k}}\isdef\m{C_{k}}\m{\Phi_{k}}$.
We employ canonical Dirac quantization \citep{Dirac1982-Book}, explicated
in App.~C of Ref.~\onlinecite{Minev2020}, with the canonical commutator~$\left[\m{\hat{\Phi}_{k}},\m{\hat{Q}_{k}}\right]=\m{\hat{\Phi}_{k}}\m{\hat{Q}_{k}}^{\intercal}-\m{\hat{Q}_{k}}\m{\hat{\Phi}_{k}}^{\intercal}=i\hbar\hat{I}\m I$,
where~$\m I$ is the identity matrix of dimension~$N-r$ and~$\hat{I}$
is the identity operator on the composite-system Hilbert space. The
Hamiltonian of the composite system is 
\begin{equation}
\hat{H}_{\mathrm{full}}=\frac{1}{2}\hat{\m Q}_{\m k}^{\intercal}\m{C_{k}}^{-1}\hat{\m Q}_{\m k}+\frac{1}{2}\hat{\m{\Phi}}_{\m k}^{\intercal}\m{L_{k}}'^{-1}\hat{\m{\Phi}}_{\m k}+\sum_{j=1}^{J}\mathcal{E}_{j}\left(\hat{\Phi}_{j}\right)\;,\label{eq:H-full-composite}
\end{equation}
where~$\m{L_{k}}'^{-1}$ is the reduced effective capacitance matrix~$\m{L_{k}}^{-1}$
with all non-linear dipole inductance~$L_{j}$ contributions subtracted
out. Since we included the non-linear dipole fluxes~$\Phi_{j}$ in
the basis~$\m{\Phi}_{\m k}$, this amounts to a simple subtraction
of the inductances from the diagonal of~$\m{L_{k}}^{-1}$. The inversion
of the composite-system capacitance matrix leads to the third capacitive
dressing of the subsystem parameters by the couplers and coupled
systems. 

Equation~\eqref{eq:H-full-composite} brings us to the desired composite-system
Hamiltonian~$\hat{H}_{\mathrm{full}}$ defined by Eqs.~\eqref{eq:H-full}
and~\eqref{eq:H-int}. The~$n$-th subsystem Hamiltonian is a partition
of the matrix equation given in Eq.~\eqref{eq:H-full-composite}, 
\begin{equation}
\hat{H}_{n}=\frac{1}{2}\hat{\m Q}_{n}^{\intercal}\m C_{n}^{-1}\hat{\m Q}_{n}+\frac{1}{2}\hat{\m{\Phi}}_{n}^{\intercal}\m L_{n}'^{-1}\hat{\m{\Phi}}_{n}+\sum_{j\in\mathcal{J}_{n}}\mathcal{E}_{j}\left(\hat{\Phi}_{j}\right)\;,\label{eq:Hn-sys}
\end{equation}
where~$\mathcal{J}_{n}$ is the set of non-linear dipoles belonging
to the~$n$-th subsystem. The total number of junctions is~$J=\sum_{n=0}^{K}\left|\mathcal{J}_{n}\right|$.
The vectors of subsystem charge and flux operators are
\begin{equation}
\hat{\m Q}_{n}\isdef\block\left(\hat{\m Q}_{\m k},\mathcal{N}_{n}\right)\quad\text{and}\quad\hat{\m{\Phi}}_{n}\isdef\block\left(\hat{\m{\Phi}}_{\m k},\mathcal{N}_{n}\right)\;,
\end{equation}
where the~$\block$ function yields the matrix partition of its first
argument with respect to the set of nodes indicated in its second
argument. Similarly, the~$n$-th subsystem matrices are the matrix
partitions with respect to the set of nodes~$\mathcal{N}_{n}$ of
the~$n$-th system,
\begin{subequations}
\label{eq:LC-for-subsystem}
\begin{align}
\m C_{n}^{-1} & \isdef\block\left(\m{C_{k}}^{-1},\mathcal{N}_{n},\mathcal{N}_{n}\right)\;,\\
\m L_{n}'^{-1} & \isdef\block\left(\m L_{n}'^{-1},\mathcal{N}_{n},\mathcal{N}_{n}\right)\;,
\end{align}
\end{subequations}
where the second and third argument indicates the set of nodes associated
with the matrix rows and columns, respectively. The~$K\left(K-1\right)/2$
pair-wise interaction terms among the~$K+1$ subsystems take the
form
\begin{equation}
\hat{H}_{n'm'}=\sum_{n\in\mathcal{N}_{n'}}\sum_{m\in\mathcal{N}_{m'}}\left(\hat{Q}_{n}\hat{Q}_{m}/\text{\ensuremath{C_{nm}^{\mathrm{eff}}}}+\hat{\Phi}_{n}\hat{\Phi}_{m}/\text{\ensuremath{L_{nm}^{\mathrm{eff}}}}\right)\;,\label{eq:H-nm-explicit}
\end{equation}
where the indices~$n'$ and~$m'$ label the subsystems, for~$n'\neq m'$.
The effective coupling capacitance~$\text{\ensuremath{C_{nm}^{\mathrm{eff}}}}$
and inductance~$\text{\ensuremath{L_{nm}^{\mathrm{eff}}}}$ associated
with the $n$-th and~$m$-th node of the~$n'$-th and~$m'$-th
subsystems, respectively, are obtained from the off-diagonal blocks
of the dressed system matrices, 
\begin{subequations}
\label{eq:Ceff-Leff-block}
\begin{align}
1/\text{\ensuremath{C_{nm}^{\mathrm{eff}}}} & \isdef2\block\left(\m{C_{k}}^{-1},\mathcal{N}_{n'}\left(n\right),\mathcal{N}_{m'}\left(m\right)\right)\;,\\
1/\text{\text{\ensuremath{L_{nm}^{\mathrm{eff}}}}} & \isdef2\block\left(\m{L_{k}}'^{-1},\mathcal{N}_{n'}\left(n\right),\mathcal{N}_{m'}\left(m\right)\right)\;,
\end{align}
\end{subequations}
where~$\mathcal{N}_{n'}\left(n\right)$ and~$\mathcal{N}_{m'}\left(m\right)$
denote the single-element subsets of~$\mathcal{N}_{n'}$ and~$\mathcal{N}_{m'}$
that comprise just the~$n$-th and~$m$-th node of the sets, respectively.

By construction, we have assumed that the subsystem Hamiltonians~$\hat{H}_{n}$
are of familiar, known structure and can be diagonalized by known
analytical or numerical methods. For instance, each Hamiltonian can
be expressed in a harmonic-oscillator basis and fully diagonalized,
as detailed in the methodology of the energy-participation-ratio method
\citep{Minev2020}. A diagonalization in this second quantized basis
can provide the~$n$-th system Hamiltonian in the diagonal form
\begin{equation}
\hat{H}_{n}=\sum_{\mu=0}^{M_{n}}\sum_{\nu=0}^{\infty}E_{n\mu\nu}\kb{n,\mu,\nu}{n,\mu,\nu}\;,
\end{equation}
where~$E_{n\mu\nu}$ is the~$\nu$-th Hamiltonian energy level of the~$\mu$-th resonant eigenmode of the~$n$-th subsystem, which
supports a total of~$M_{n}$ number of quantized resonant modes.
Since we have partitioned the composite into small, manageable subsystems,
for an effectively lumped subsystem of interest one can diagonalize~$\hat{H}_{n}$
independently from the circuit using analytical or numerical techniques.
In the case of numerics, one can use a large, truncated basis for
the diagonalization, but subsequently can retain only a handful of
the low-energy levels. The flux~$\hat{\Phi}_{n}$ and charge~$\hat{Q}_{n}$
 operators can be constructed in this truncated eigenbasis; i.e.,
in terms of the~$\ket{n,\mu_{n},\nu_{n}}$ and~$\ket{m,\mu_{m},\nu_{m}}$
eigenstates. Thus, the interaction~$\hat{H}_{nm}$ and composite
Hamiltonian~$\hat{H}_{\mathrm{full}}$ can be expressed more efficiently
\footnote{This is the form employed in a recent package entitled \href{https://scqubits.readthedocs.io/en/latest/index.html}{scqubits};
see also Ref.~\onlinecite{Kerman2020}.}.

In this section, we explicitly constructed the subsystem Hamiltonians
and their interactions and eliminated all coupler degrees of freedom and
associated constraints, and incorporated all circuit effects, such
as the~$A^{2}$ diamagnetic contribution, in Eqs.~\eqref{eq:H-full-composite}\textendash \eqref{eq:Ceff-Leff-block}.
The parameters of each subsystem Hamiltonian~$\hat{H}_{n}$ were
obtained by three sequential dressings. Thus, the spectrum of~$\hat{H}_{n}$
and~$\hat{H}_{m}$ are not circuit parameter independent in general.
They can be dressed by their couplers and by each other's circuit
parameters. The dressing can in principle be rather non-local. This
depends on the particulars of the circuits. In typical situations,
nodes in the network graph separated by a minimal path of several
edges can be coupled but potentially negligibly so. For example, in
the inversion step of~$\m{C_{k}}$, these distance direct couplings
are allowed but severely dampened by the weight of the inversion.
However, dressing of node parameters by proximal coupler edges can
be large. For example, for the direct capacitance between the pads
of a transmon qubit can be renormalized by a factor of two subject
to a large pad coupler, such as featured in our experiment, see Sec.~\ref{sec:Experiment}.
The coupling terms are also similarly dressed by the system parameters.
Proper elimination at the composite level, rather than at the individual-system
level, is essential for more general circuits and accurate results,
as observed from data for the theory versus experiment comparison
presented in Sec.~\ref{sec:Experiment}. 

\subsection{Example of the experimentally measured devices}

Let us briefly illustrate the method by returning to the example of
Fig.~\ref{fig:model}. The composite system is comprised of the qubit ($\mathcal{N}_{0}=\left\{ P_{1}\right\} $),
CPW readout resonator~($\mathcal{N}_{1}=\left\{ B_{1},\ldots\right\} $),
two CPW bus resonators~($\mathcal{N}_{2}=\left\{ B_{2},\ldots\right\} $
and~$\mathcal{N}_{3}=\left\{ B_{3},\ldots\right\} $), and coupler
nodes~($\mathcal{N}_{\mathrm{coupler}}=\left\{ P_{0},C\right\} $);
the ellipses in the sets indicate the continuum of nodes associated
with the distributed lines. Here,~$K=3$; however, we only need one
lumped simulation cell. This cell contains the set of nodes~$\mathcal{N}_{0}^{\mathrm{cell}}=\left\{ P_{0},P_{1},B_{0},B_{1},B_{2},C\right\} $;
the ground node is considered accessible by every cell. The remaining
three cells associated with the bodies of the CPW transmission lines
need not be simulated explicitly since they can be handled quasi-analytically,
see Sec.~\ref{sec:LTL}. Following the procedure outlined in Sec.~\ref{sec:Lumped-cell-model},
the cell capacitance matrix~$\m C_{0,\mathrm{cell}}$ is extracted;
its fully-connected graph is depicted with green lines in Fig.~\ref{fig:model}(c).
For our experiment, we employed the \emph{Ansys~Q3D Extractor} quasi-field
solver. The creation of the qubit cell and extraction of its Maxwell
matrix was automated using \textsc{Qiskit Metal} \citep{Qiskit_Metal}. 

The inductance matrix of the qubit cell contains a single inductor
due to the one junction in the cell; i.e., $J=1$, $\mathcal{J}_{0}=\left\{ \Phi_{J}\right\} $
and~$\mathcal{J}_{n}=\emptyset$ for~$n>0$, where the junction
flux is~$\Phi_{J}=\Phi_{P_{1}}-\Phi_{P_{0}}$. The CPW cells all
have diagonal capacitance matrices~$\m C_{n,\mathrm{cell}}$, where
$n\in\left\{ 1,2,3\right\} $; their inductance matrices~$\m L_{n,\mathrm{cell}}^{-1}$
each have a one-dimensional kernel space. The qubit cell inductance
matrix in isolation has~$6$ columns but has unity rank. Since the
three bus nodes are touched by inductors in other systems, these three
nodes will not be eliminated; i.e., one should consider the kernel
space of the composite inductance matrix~$\m L_{\m n}^{-1}$ and
not just that of a given cell~$\m L_{n,\mathrm{cell}}^{-1}$, see
Eq.~\eqref{eq:CL-composition}. 

Since there are no coupling inductors, we can obtain the reduced
Hamiltonian matrices~$\m{C_{k}}^{-1}$ and~$\m{L_{k}}^{-1}$ directly
from Eq.~\eqref{eq:reduced-Li-C}, without having to perform a second
elimination of inductive coupling nodes. Since the transmission line
systems have diagonal node-to-datum cell matrices, it suffices to
only invert and reduce~$\m C_{0,\mathrm{cell}}$ to obtain all needed
information comprised in~$\m{C_{k}}^{-1}$. The reduced effective
circuit is depicted in Fig.~\ref{fig:model}(d); the third bus is
omitted from the illustration for the sake of visual simplicity. The
dressed transmon qubit Hamiltonian, associated circuit shaded in green
in Fig.~\ref{fig:model}(d), is 
\[
\hat{H}_{0}=\frac{1}{2}\left(\hat{Q}_{J}-Q_{\mathrm{ofs}}\right)^{2}/C_{J}^{\mathrm{eff}}-E_{J}\cos\left(\hat{\Phi}_{J}\right)\;,
\]
where we have included a potential charge offset~$Q_{\mathrm{ofs}}$,
which could be due to a charge fluctuation on any of the nodes in~$\mathcal{N}_{0}^{\mathrm{cell}}$,
where~$\hat{Q}_{J}\isdef C_{J}^{\mathrm{eff}}\ddt\hat{\Phi}_{J}$,
and~$1/C_{J}^{\mathrm{eff}}\isdef\block\left(\m{C_{k}}^{-1},\left\{ P_{1}\right\} ,\left\{ P_{1}\right\} \right)$.
Importantly,~$C_{J}^{\mathrm{eff}}$ comprises the dressed and renormalization
effect of all capacitances in the qubit cell model; i.e., while we
may refer to it as the dressed qubit capacitance, its value depends
on the coupling and transmission-line end-loading capacitances. That
is, the qubit and the neighboring system Hamiltonians are dressed
by each others' elements and their parameters are not strictly independent.
This first dressing is purely classical and due to linear effects.
The dressed loading capacitance of the~$n$-th line, see Eq.~\eqref{eq:LTL-H-initial-qm},
is 
\[
1/C_{L,n}^{\mathrm{eff}}\isdef\block\left(\m{C_{k}}^{-1},\left\{ B_{n}\right\} ,\left\{ B_{n}\right\} \right)\;.
\]
The interaction Hamiltonian between the $n$-th and $m$-th subsystems is
\[
\hat{H}_{nm}=\hat{Q}_{n}\hat{Q}_{m}/\text{\ensuremath{C_{nm}^{\mathrm{eff}}}}\;,
\]
where~$\hat{Q}_{0}=\hat{Q}_{J}$ and~$\text{\ensuremath{C_{nm}^{\mathrm{eff}}}}$
is the appropriately scaled off-diagonal term of the reduced and inverted~$\m C_{0,\mathrm{cell}}$,
see Eq.~\eqref{eq:Ceff-Leff-block}. In Sec.~\ref{sec:Experiment}, we compare the parameters of interest found from this method to those experimentally measured on a large-scale quantum processor \citep{Corcoles2021}.

\subsection{Lumped cell model}

\label{sec:Lumped-cell-model}

In the physical model of a lumped cell, see for example Fig.~\ref{fig:model}(b),
each galvanically-isolated conducting island is represented by a node
in the circuit. The effect of the geometry and all other physical
aspects of the cell, such as dielectric permittivities and material
properties, on the electrostatics of the cell are succinctly expressed
in the Maxwell capacitance matrix\citep{zangwill2012-book}~$\m{C_{M}}$,
which is reviewed in App.~\ref{subsec:Maxwell-Capacitance-Matrix}.
The~$\left(i,j\right)$-th off-diagonal element of the Maxwell matrix
is the mutual capacitance between the~$i$-th and~$j$-th node in
the cell times negative unity.  It is efficiently extracted from
the cell model using a quasi-field solver, such as \emph{Ansys~Q3D
Extractor}; this process\emph{ }is automated\emph{ }in our open-source
project \textsc{Qiskit Metal} \citep{Note1}\emph{.} The cell capacitance
matrix with respect to the cell node-to-datum generalized magnetic
fluxes, see Eqs.~\eqref{eq:CL-composition} and~\eqref{eq:Cmaxwell},
is
\begin{equation}
\m C_{\mathrm{cell}}=\m S_{\m M}^{\intercal}\m{C_{M}}\m{S_{M}}\;,\label{eq:Maxwell-mat-reduced}
\end{equation}
where~$\m{S_{M}}\isdef\left[\begin{array}{cc}
\m 0_{N} & \m I_{N}\end{array}\right]^{\intercal}$,~$\m 0_{N}$ is the zero column vector of length~$N$, and~$\m I_{N}$
is the identity matrix of dimension~$N$, where~$N=\dim\m{C_{M}}-1$.

\section{Quantum physics of the loaded transmission-line cell}

\label{sec:LTL}

\paragraph{}

\begin{figure}
\begin{centering}
\includegraphics{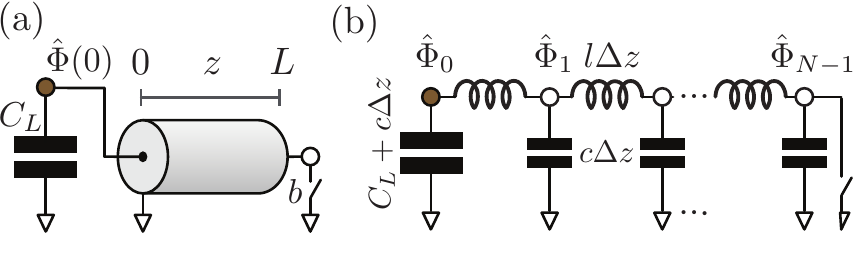}
\par\end{centering}
\caption{\label{fig:ltl-model}Transmission line capacitively loaded at one end.
(a) Continuous model of a line with total length~$L$ and loading capacitance~$C_{L}$.
Length along the line is~$z$.
For $b=0$ (resp., $b=1$), the right end of the line is terminated
in an open (resp., short). 
(b) Circuit schematic of an equivalent,
discrete-space model, with~$N$ nodes. 
The spatial discretization length is~$\Delta z\protect\isdef L/(N-1)$. 
The capacitance and inductance per unit length of the line are~$c$ and~$l$, respectively. 
The operators of the node-to-datum generalized magnetic fluxes of the
circuit are~$\hat{\Phi}_{n}$, for~$n=0,1,\ldots N-1$.}
\end{figure}

Because of its practical and central importance, we revisit the diagonalization
of an end-loaded transmission line, depicted in Fig.~\ref{fig:ltl-model}(a).
Quantizing its discrete model and taking the continuous limit, one
finds its quantum Hamiltonian
\begin{multline}
\hat{H}_{\mathrm{LTL}}=\frac{1}{2}C_{L}^{-1}\left[\hat{Q}(0)\right]^{2}+\int_{0^{+}}^{L}\mathrm{d}z\,c^{-1}\left[\hat{q}\left(z\right)\right]^{2}\\
+\int_{0}^{L}\mathrm{d}z\,l^{-1}\left[\frac{\partial\hat{\Phi}\left(z\right)}{\partial z}\right]^{2}\;,\label{eq:LTL-H-initial-qm}
\end{multline}
where $C_{L}$ is the effective end-loading capacitance to ground
at $z=0$, the total charge operator at $z=0$ is $\hat{Q}(0)$, and
$L,c,l,\hat{q}\left(z\right),$ and $\hat{\Phi}\left(z\right)$ are
the line total length, capacitance and inductance per unit length,
and the charge-density and magnetic-flux-field operators at length
$z$ along the line, respectively. Due to the end-singularity in the
capacitance per unit length of the line, strictly speaking, there
is no operator $\hat{q}\left(0\right)$; hence, the integration from
$0^{+}$ in the lower integral bound, which excludes zero. The scalar,
one-dimensional quantum fields obey $\left[\hat{\Phi}(z),\hat{q}(z')\right]=i\hbar\delta\left(z-z'\right)\hat{I}$
and $\left[\frac{\partial\hat{\Phi}\left(z\right)}{\partial z},\hat{q}(z')\right]=i\hbar\frac{\partial}{\partial z}\delta\left(z-z'\right)\hat{I}$,
where $\hat{I}$ is the identity operator.

\paragraph{Boundary conditions.}

For an open (short) right-end termination, the right boundary condition
is of the standard Dirichlet (respectively, Neumann) type, $\frac{\partial\hat{\Phi}\left(L\right)}{\partial z}=\hat{0}$
(resp., $\hat{\Phi}\left(L\right)=\hat{0}$). However, the left-end
boundary condition is non-standard. It is not of the type covered
by standard Sturm\textendash Liouville theory, $\frac{\partial\hat{\Phi}\left(0\right)}{\partial z}=lC_{L}\frac{\partial^{2}\hat{\Phi}\left(0\right)}{\partial t^{2}}$.
This condition makes the boundary condition eigenvalue-dependent leading
to transcendental eigensolutions. The equation of motion in the interior
of the line is the standard wave equation, $v_{p}^{2}\frac{\partial^{2}\hat{\Phi}}{\partial z^{2}}\left(z\right)=\frac{\partial^{2}\hat{\Phi}}{\partial t^{2}}\left(z\right),$
valid on $0<z<L$, where the phase velocity is $v_{p}=\sqrt{1/\left(lc\right)}$.

\paragraph{Characteristic eigenvalue equation.}

The Hamiltonian is diagonalized using harmonic solutions and the superposition
principle for linear systems. It is easier to show by demoting the
problem to classical fields, then recovering the quantum field operators
and their quantum zero-point fluctuations using the energy-participation-ratio
method \citep{Minev2020}; this establishes the classical-quantum
correspondence of the problem and avoids technical detail regarding
the proper orthonormalization of the field eigensolution in the presence
of the loading irregularity. Using the harmonic ansatz $\Phi_{m}\left(z,t\right)=u_{m}\left(z\right)\Phi_{m}\left(t\right)=A_{m}\sin\left(\omega_{m}t\right)\cos\left(k_{m}z+\phi_{m}\right)$ and
eliminating~$k_{m}$ and~$\phi_{m}$, we find the characteristic
eigenvalue equation for $\omega_{m}$,
\begin{equation}
\boxed{\omega_{m}\frac{L}{v_{p}}+\arctan\left(\frac{\omega_{m}}{\omega_{L}}\right)=m\pi+b\frac{\pi}{2}:m\in\mathbb{Z}_{\geq0}\;,}\label{eq:LTL-char-eq}
\end{equation}
where $b=0$ (resp., $b=1)$ for an open (resp., short) right-end
boundary condition and $\omega_{L}\isdef\frac{1}{lC_{L}v_{p}}=\frac{1}{C_{L}Z_{0}}$.
The non-negative integer~$m$ labels the discrete, uncountable number
of modes. The associated spatial phase shift and wavenumber are $k_{m}=\omega_{m}/v_{p}$
and $\phi_{m}=\arctan\left(\omega_{m}/\omega_{L}\right)$, respectively. 

\begin{figure}
\begin{centering}
\includegraphics{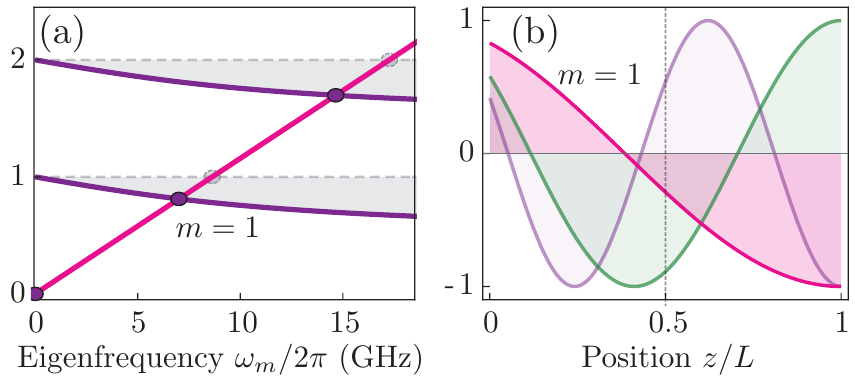}
\par\end{centering}
\caption{\label{fig:Eigenmodes-of-a}Dressing of the eigenfrequencies~(a) and eigenfields~(b)
of a transmission line resonator capacitively loaded at its left end.
Line and loading parameters correspond to those of measured devices. 
(a) Solutions~$\omega_{m}$ of the transcendental eigenvalue equation are found at the intersection of the  diagonal (red) line with the curved (resp., horizontal) line for the loaded (resp., unloaded) line, see Eq.~\eqref{eq:LTL-char-eq}. 
The unloaded line corresponds to a loading capacitance~$C_{L}=0$.
The point labeled~$m=1$ denotes the fundamental loaded solution.
(b) Corresponding eigenfield spatial distributions~$u_{m}\left(z\right)$, depicted for the first three modes, indexed by~$m$. Vertical dashed line is a guide to the eye, highlighting the symmetry breaking due to the asymmetric pull on the fields toward the loading.}
\end{figure}

The equation is transcendental. For a loading knee frequency $\omega_{L}\gg\omega_{m}$
(resp., $\omega_{L}\ll\omega_{m}$) the loading capacitor acts as
an open (resp., short), while for $\omega_{m}$ of the same order
of magnitude as $\omega_{L}$, the loading exerts a significant renormalization
on the eigenfrequency and eigenfields away from those obtainable from
the unloaded transmission line solutions; see Fig. \ref{fig:Eigenmodes-of-a}.
In other words, in this regime, the naive eigenfunctions of an unloaded
transmission line are not a good starting-point basis choice, since
the phase shift $\phi_{m}$ and renormalization of the mode frequencies
can be large. It is this latter regime that is experimentally relevant
for the devices and measurements presented in Sec. \ref{sec:Experiment}.
For our devices, the renormalization of the loading was on the order
of 20\textendash 25\%.

\paragraph{Quantum Hamiltonian.}

Exploiting the eigensolutions~$\omega_{m}$ of Eq. \eqref{eq:LTL-char-eq},
in second quantization, Eq. \eqref{eq:LTL-H-initial-qm} becomes
\[
\hat{H}_{\mathrm{LTL}}=\sum_{m=m_{0}}^{\infty}\hbar\omega_{m}\left(\hat{a}_{m}^{\dagger}\hat{a}_{m}+\frac{1}{2}\right)\;,
\]
where~$\hat{a}_{m}$ is the annihilation operator for the $m$-th
mode. For the case of an open termination~$b=0$, a trivial zero-frequency
(d.c.) mode solution of Eq. \eqref{eq:LTL-char-eq} exists, and the
lower bound of the sum is~$m_{0}=1$; otherwise, for~$b=1$,~$m_{0}=0$.
By linearity, the fields are linear superpositions of the modal operators,
\begin{subequations}
\label{eq:LTL-fields}
\begin{align}
\hat{\Phi}\left(z\right) & =\sum_{m=m_{0}}^{\infty}\Phi_{m}^{\mathrm{ZPF}}\left(z\right)\left(\hat{a}_{m}^{\dagger}+\hat{a}_{m}\right)\;,\nonumber \\
\hat{q}\left(z\right) & =\sum_{m=m_{0}}^{\infty}iq_{m}^{\mathrm{ZPF}}\left(z\right)\left(\hat{a}_{m}^{\dagger}-\hat{a}_{m}\right)\;,\nonumber \\
\hat{Q}\left(0\right) & =\sum_{m=m_{0}}^{\infty}iQ_{m}^{\mathrm{ZPF}}\left(0\right)\left(\hat{a}_{m}^{\dagger}-\hat{a}_{m}\right)\;,\label{eq:LTL-fields-Q0}
\end{align}
\end{subequations}
where we will use the EPR to find the values of the quantum zero-point
fluctuations $\Phi_{m}^{\mathrm{ZPF}}\left(z\right)$, $q_{m}^{\mathrm{ZPF}}\left(z\right)$,
and $Q_{m}^{\mathrm{ZPF}}\left(0\right)$ of the magnetic flux, charge
density, and charge, respectively.

\paragraph{Quantizing the fields using the energy-participation ratio (EPR). }

The EPR of the loading capacitor $p_{mL}$ in mode $m$ is the fraction
of energy stored in the loading capacitor $\mathcal{E}_{C_{L}}$ relative
to the total capacitive energy of the mode $\mathcal{E}_{\mathrm{cap}}$.
As a fraction, the participation is independent of the normalization
of $u_{m}$; in the classical setting,
\begin{equation}
p_{mL}\isdef\frac{\mathcal{E}_{C_{L}}}{\mathcal{E}_{\mathrm{cap}}}=\frac{\frac{1}{2}C_{L}\left[u_{m}\left(0\right)\right]^{2}}{\frac{1}{2}C_{L}\left[u_{m}\left(0\right)\right]^{2}+\frac{1}{2}\int_{0}^{L}\mathrm{d}z\,c\left[u_{m}\left(z\right)\right]^{2}}\;,\label{eq:LTL-epr-cm}
\end{equation}
which is evaluated using the solutions of Eq. \eqref{eq:LTL-char-eq}. 

In the quantum setting, $\left\langle \mathcal{E}_{C_{L}}\right\rangle =\left\langle \frac{1}{2C_{L}}\left[\hat{Q}\left(0\right)\right]^{2}\right\rangle $,
where we take the expectation value over a single photon state in
mode $m$ and disregarding zero-point energy contributions; substituting
Eq. \eqref{eq:LTL-fields-Q0}, $\left\langle \mathcal{E}_{C_{L}}\right\rangle =\frac{1}{C_{L}}Q_{\mathrm{ZPF}}^{2}$.
The total capacitive energy is $\left\langle \hat{\mathcal{E}}_{\mathrm{cap}}\right\rangle =\frac{1}{2}\hbar\omega_{m}$;
hence, the quantum ZPF of the field charge operator at $z=0$
can be found in terms of the EPR, which we can calculate from the
classical solutions, using Eq. \eqref{eq:LTL-epr-cm},
\begin{equation}
\left[Q_{m}^{\mathrm{ZPF}}\left(0\right)\right]^{2}=\text{\ensuremath{\frac{\hbar\omega_{m}}{2}C_{L}p_{mL}}}\;.\label{eq:LTL-QZPF-0}
\end{equation}
Using the same line of EPR reasoning, one finds 
\begin{align}
\left[q_{m}^{\mathrm{ZPF}}\left(z\right)\right]^{2} & =\frac{\hbar\omega_{m}}{2}cp_{mc}\left(z\right)\;,\label{eq:LTL-qzpf-andPhi_ZPF}\\
\Phi_{m}^{\mathrm{ZPF}}\left(z\right) & =\frac{1}{c\omega_{m}}q_{m}^{\mathrm{ZPF}}\left(z\right)\;,
\end{align}
where the EPR density~$p_{mc}\left(z\right)\isdef\frac{1}{2}c\left[u_{m}\left(z\right)\right]^{2}/\mathcal{E}_{\mathrm{cap}}$
is the density of the fraction of capacitive energy stored in the
infinitesimal capacitance at~$z$. Equations \eqref{eq:LTL-QZPF-0}
and~\eqref{eq:LTL-qzpf-andPhi_ZPF} fully specify the quantum fields,
Eq. \eqref{eq:LTL-fields}, and thus complete the solution of the
loaded line.

\section{Measurements and comparison between theory and experiment}

\label{sec:Experiment}

\begin{figure}
\begin{centering}
\includegraphics{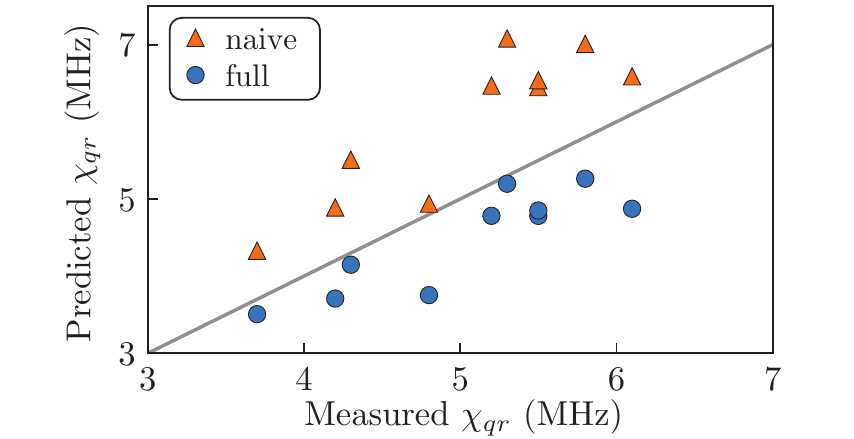}
\par\end{centering}
\caption{\label{fig:expeirment}Measured vs. predicted qubit-readout cross-Kerr coupling~$\chi_{qr}$ for 10 qubit subsystems
across two different 14-qubit processors. 
The parameter~$\chi_{qr}$ is the most sensitive and dressed parameter of the composite-system Hamiltonian. The frequency of the readout CPW resonator is 
dressed down by~25\% due to its embedding in the larger network.
Circles: predictions from the full method of this paper.
Triangles: predictions from a simpler model that performs an approximate capacitive reduction based on weak coupling and does not account for the dressing of the zero-point quantum fluctuations of the CPWs. The agreement between the experiment and the full (resp., naive)
theory is~-10.5\% (resp., +19\%). The simulation error bars are smaller
than the size of the dots.}
\end{figure}

{
	\setlength{\extrarowheight}{3pt}
	\setlength{\doublerulesep}{4pt}
	\addtolength{\tabcolsep}{10pt}
	\rowcolors{1}{}{gray!4}
\begin{table*}
\begin{centering}
\begin{tabular}{ll}
\textbf{Model feature / Parameter}  & \textbf{Magnitude of effect on experimental agreement}\tabularnewline
\hline 
Including all coupling Hamiltonians~$\hat{H}_{nm}$ & $5\%$\tabularnewline
Including qubit coupling to all bus resonators~$\hat{H}_{qm}$ & $1\%$\tabularnewline
Transmission-line impedance~$Z_{0}$ & $1\%$ for $3\%$ variation on~$Z_{0}=50\,\Omega$\tabularnewline
Chip separation & $1\%$ for $20\%$ variation on separation\tabularnewline
Substrate permittivity~$\epsilon_{r}$ & $0.5\%$ for $2\%$ variation on~$\epsilon_{r}$\tabularnewline
Including readout first harmonic in analysis~$\omega_{12}$ & $0.3\%$\tabularnewline
Cell partition bounding box length & Less than~$0.5\%$ and increasingly negligible over 100~$\mathrm{\mu m}$\tabularnewline
Bus resonator frequency~$\omega_{n1}$ & Negligible for $\pm5\%$ variation on~$\omega_{n1}$\tabularnewline
Cell bounding box padding distance & Negligible over 100~$\mathrm{\mu m}$\tabularnewline
Transmission-line phase velocity~$v_{p}$ & Negligible\tabularnewline
Including charge line node in reduced model & Negligible\tabularnewline
Sample holder enclosure distance & Negligible\tabularnewline
\end{tabular}
\par\end{centering}
\caption{\label{tab:Parameter-variation-budget}Influence and error budget
for the experimental agreement due to model features and 
parameters. Left column: model feature or parameter. Right column:
magnitude of the change in the average experimental agreement reported
in Fig.~\ref{fig:expeirment} for the dispersive coupling~$\chi_{qr}$.
The intrinsic capacitance~$C_{j}$ of a Josephson junction in the
model is fixed, and its inductance~$L_{J}$ is determined based on
the measured qubit frequency.
}
\end{table*}
}

We applied the lumped-oscillator-model~(LOM) method, presented in
this work, on two 14-qubit superconducting quantum processors, measured
over multiple cooldowns\citep{Corcoles2021}. Ten of the most well measured subsystems were chosen for analysis with the LOM method. The processor architecture
was based on floating transmon qubits \citep{Koch2007} interconnected
by CPW bus resonators \citep{Blais2004}, controlled by charge lines,
and readout by CPW resonators. Qubit, bus, and resonator frequencies
were allocated in the 5.1\textendash 5.5~GHz, 6.5\textendash 6.7~GHz,
and 7\textendash 7.1~GHz bands, respectively. Qubit anharmonicities
were measured in the 300\textendash 350~MHz range. Readout coupling
was strong so as to provide fast qubit readout \citep{Corcoles2021};
qubit-cavity dispersive shifts~$\chi_{qr}$ were in the 3\textendash 7~MHz
range. Each qubit had either 3 or 4 couplers\textemdash one coupler
for the readout CPW resonator and the remainder for bus
CPW resonators; see Fig.~\ref{fig:model}(b). The qubit connectivity
implemented the topology of 3 lattice placates. FPGAs controlled the
experiment. Single-qubit gates were Gaussian-shaped pulses of length
30~ns. The measurement readout time was 320~ns, performed in reflection,
through a Traveling Wave Parametric Amplifier (TWPA). The experimental is further described in~\onlinecite{Corcoles2021}. 

We measured the qubit and cavity system parameters\textemdash the
qubit frequency~$\omega_{q}$ and anharmonicity~$\alpha_{q}$ and
the readout frequency~$\omega_{r}$\textemdash using standard spectroscopic
and time-resolved protocols. Of note, we used a more-sensitive
Ramsey dressed-dephasing-based method \citep{Gambetta2006,Gambetta2008-qm-traj,Corcoles2021}
to measure their interaction~$\chi_{qr}$. This much smaller parameter~$\chi_{qr}$
was extracted from dephasing experiment measurements simultaneously fitting the dephasing in both the~$\left\langle X\right\rangle $ and~$\left\langle Y\right\rangle $
quadratures of the qubit Bloch vector as a function of the readout probe frequency and the effective readout-strength photon number~$\bar{n}$.
This collection of curves, which are very sensitive to the Hamiltonian
parameters, was simultaneously fit to extract the readout photon number~$\bar{n}$
and dispersive shift~$\chi_{qr}$. We also accounted for the rise
and fall times of the readout pulse in the extraction by accounting
for the time-domain evolution of the reduced stochastic master equation
of the qubit \citep{Gambetta2008-qm-traj}.

In the following, we compare the measured Hamiltonian parameters to
those obtained using the LOM method applied to the physical layout
of the devices. We focus on the most challenging parameter to obtain
quantitative agreement for,~$\chi_{qr}$.

We model the qubit cell as depicted in Fig.~\ref{fig:model}(b)
by including; a short segment of charge control line, neighboring
CPW structures, and all of the coupler structures attached to the qubit
in the cell model. We extract the qubit cell Maxwell capacitance matrix~$\m{C_{M}}$
using \emph{Ansys Q3D Extractor} by using the layout geometry and
a nominal substrate relative permittivity \citep{Krupka2006}~$\epsilon_{r}=11.45$.
We estimate the nominal Josephson junction intrinsic capacitance~$C_{J}=2\,\mathrm{fF}$.
The matrix is treated and reduced as described in Sec.~\ref{sec:cell-model},
after which we find a notably large capacitance to ground $C_{L}$,
in excess of $320\,\mathrm{fF}$, loading the readout coupling node
$P_{0}$. The effective CPW lengths are extracted from Eq.~\eqref{eq:LTL-char-eq}, where 
 impedance~$Z_{0}=53\,\Omega$ and phase
velocity~$v_{p}=0.403c$, where $c$ is the speed of light, of the
lines were found analytically from the line geometry \citep{Simons2001}.
Finally, the qubit Josephson tunnel junction inductance~$L_{j}$
is varied from its nominal values, inferred from room-temperature
resistance measurements~\citep{Gloos2000}, due to its unavoidable
and inherent aging and cool-down variability, to obtain a qubit frequency
agreement at the 1\% level.

While some parameters of the model are known with exceedingly high
precision, such as the geometry of the qubit device, we allow for
small, reasonable variation in several parameters that are less well
known. The variation in~$\chi_{qr}$ caused by the roughly~5\% uncertainty
inherent in these parameters, including $L_{J},C_{J},Z_{0}$ and $\epsilon_{r}$,
is detailed in Table~\ref{tab:Parameter-variation-budget}. Furthermore,
for achieving the highest level of agreement, we detail in Table~\ref{tab:Parameter-variation-budget}
the relative importance of including finer-order effects in the Hamiltonian
model, such as the coupling of the qubit to the readout CPW first-harmonic
mode. While this particular contribution has only a single-percent
level effect on~$\chi_{qr}$, including all direct CPW-CPW couplings
originating from the direct capacitive links in the qubit cell, described
in the Hamiltonian as terms of the form~$\beta_{nm}\hat{a}_{n}^{\dagger}\hat{a}_{n}\hat{a}_{m}^{\dagger}\hat{a}_{m}$,
significantly decreases the cross-Kerr~$\chi_{qr}$ by approximately~5\%.
In practice, since the qubit cell is only strongly coupled to a small handful
of modes, we numerically diagonalize each individual
system Hamiltonian~$\hat{H}_{n}$ first, constructing the
composite Hamiltonian~$\hat{H}_{\mathrm{full}}$, using Eqs.~\eqref{eq:H-full}
and~\eqref{eq:H-int} and diagonalizing the full system, from which
we extract all final model parameters.

Accounting for these finer effects, we present the comparison between
experimentally measured and LOM-predicted values of~$\chi_{qr}$ in
Fig.~\ref{fig:expeirment}. Across the~10 devices, we find that
the agreement between theory and experiment is~-10.5\%. We compare
this to a naive model, which disregards these finer effects and, chiefly,
does not account for the large renormalization of the transmission
line eigenmodes as detailed in Sec.~\ref{sec:LTL}, illustrated in
Fig.~\ref{fig:Eigenmodes-of-a}. Under these more conventional approximations,
the average theory-experiment agreement is +19\%. A chief component
in the difference is the large magnitude of~$C_{L}$, which dresses
down the CPW readout mode frequency from approximately~8.8~GHz to
7.0~GHz, a 25\% renormalization effect. 

In Table~\ref{tab:Parameter-variation-budget}, we summarize the fluctuation in the agreement between simulated and measured~$\chi_{qr}$ from the modification of model features and parameters . For example, the table reports the effect of varying
the location of the qubit cell partition and the length of neighboring
CPWs included in the qubit cell model. This is found to have a nearly negligible effect. Determination of the consistently correct amount to include was not reached, though for the transmon layouts this paper considered, $\approx~100\mu$m proved to be sound. The separation of the device chips has a small
effect when the gap is larger than the feature dimensions \citep{Minev2016}, though becomes significant when said gap is equivalent to or less than feature dimensions. 
Including the sample holder packaging in the model has a negligible
effect for the dimensions of our device. It is not possible to achieve
perfect experimental agreement by varying the model parameters within
experimental tolerances. In fact, the model is fairly well constrained.
We believe that better agreement requires a more complete model description
of distributed effects. These can be captured by more computationally-expensive,
but also more informationally-complete treatments, such as impedance
\citep{Nigg2012} or energy-participation ratio (EPR) quantization
\citep{Minev2020}. 

\paragraph{Conclusion.}

We introduced the lumped-oscillator method (LOM) and studied its  experimental performance on state-of-the-art superconducting quantum processors \citep{Corcoles2021}.
We found the LOM method to be practical, due to its systematic and modular analysis flow. It operates by partitioning the device in a
two-fold manner\textemdash subsystems and simulation cells. Cells are  small domains of the physical device layout, which can be independently analyzed using analytical results or numerical simulations. Breaking up the device layout into independent cells increases the computational modeling efficiency. Subsystems serve as small, familiar building blocks. Within the lumped approximation,
the full Hamiltonian~$\hat{H}_{\mathrm{full}}$ of the composite system is extracted with
no approximations on the strength or nature of the inductive non-linear
dipoles. Subsystem-subsystem couplings can be arbitrarily
large. This was experimentally important to account for the data. 
We  observed
subsystem renormalizations on the order of 25\% in the experiment due to dressings induced
by the coupler structures and the subsystem embedding in the larger network. Overall, we found a two-times improvement in experimental agreement using 
the LOM method over one that makes weak coupling approximations and does not fully account for the non-perturbative dressing of the distributed modes.

With the increased complexity and demands driven by the rapid development of quantum hardware, we believe fast, accurate, and systematic techniques
such as the LOM method presented here will be an essential ingredient
in the development of current and future quantum technology. For this
reason, we contribute the automation and implementation of this method
to the community, as part of our open-source project \textsc{Qiskit
Metal | for quantum device design} \citep{Note1}.

\paragraph{Acknowledgements.}

We thank M.~Malekakhlagh, F.~Solgun, D.C.~McKay, D. Wang,
and R.~Guti\'errez-J\'auregui for valuable discussions. We are grateful
to all the early-access participants of \textsc{Qiskit Metal} and the Metal team for
discussions and stress testing the open-source code. We thank G.~Calusine
and W.~Oliver for providing the traveling-wave parametric amplifier
used in this work. We acknowledge partial support for work on the
simulations, and experimental bring-up and characterization of the
devices from the Intelligence Advanced Research Projects Activity
(IARPA) under Contract No.~W911NF-16-1-0114.

\paragraph{Author contributions. }

Z.K.M developed the theory and software, and wrote the manuscript.
Z.K.M. conceived of the project based on work by J.M.G., who oversaw
the work. Z.K.M. and T.G.M. simulated and analyzed the devices. M.T.
and A.C. performed the experimental bring-up and characterization
of the devices and worked with Z.K.M on the analysis of the experimental
data. All authors discussed the results and contributed to the manuscript.

\paragraph{Code availability. }

The source code for \textsc{Qiskit Metal | for quantum device design}
and for the LOM method implementation is open-sourced and can be found
at \href{https://github.com/Qiskit/qiskit-metal}{github.com/qiskit/qiskit-metal}.

\appendix
\global\long\def\thefigure{A\arabic{figure}}%

\setcounter{figure}{0}

\numberwithin{equation}{section} 
\numberwithin{figure}{section} 
\renewcommand\theequation{\Alph{section}\arabic{equation}}
\renewcommand\thefigure{\Alph{section}\arabic{figure}}

\section{Maxwell capacitance matrix}

\label{subsec:Maxwell-Capacitance-Matrix}

The effective capacitances of a cell can be extracted from its physical
layout. A simulation incorporating the cell geometry, materials, and
electromagnetic boundary conditions can yield its Maxwell capacitance
matrix~$\m{C_{M}}$. For a cell with~$N$ nodes (nets),~$\m{C_{M}}$
is an~$N+1$ square, full-rank matrix. Its off-diagonal entries are
the negative of the capacitance~$C_{ij}$ between nodes~$i$ and~$j$,
where~$C_{ij}\geq0$. Its~$i$-th diagonal is the self-capacitance
of the~$i$-th node to infinity~$C_{ii}$ plus the value of all
coupled capacitances; i.e., the~$i$-th row and~$j$-th column element
of~$\m{C_{M}}$ is
\begin{equation}
\left[\m{C_{M}}\right]_{i,j}=\begin{cases}
-C_{ij}\;, & i\neq j\;,\\
\sum_{j'=1}^{N}C_{ij'} & i=j\;.
\end{cases}\label{eq:Cmaxwell}
\end{equation}
For example, the Maxwell matrix for the set of nodes~$\mathcal{N}=\left\{ n_{0},\ldots n_{5}\right\} $
has the form 
\[
\m{C_{M}}={\znm\begin{pmatrix} & n_{0} & n_{1} & n_{2} & n_{3} & n_{4} & n_{5}\\
n_{0} & C_{0\Sigma} & -C_{01} & -C_{02} & -C_{03} & -C_{04} & -C_{05}\\
n_{1} &  & C_{1\Sigma} & -C_{12} & -C_{13} & -C_{14} & -C_{15}\\
n_{2} &  &  & C_{2\Sigma} & -C_{23} & -C_{24} & -C_{25}\\
n_{3} &  &  &  & C_{3\Sigma} & -C_{34} & -C_{35}\\
n_{4} &  &  &  &  & C_{4\Sigma} & -C_{45}\\
n_{5} &  &  &  &  &  & C_{5\Sigma}
\end{pmatrix}},
\]
where we introduce the shorthand~$C_{i\Sigma}\isdef\sum_{j=1}^{N}C_{ij}$
and, for simplicity, we omit the lower-triangular block of the symmetric
matrix. The sum of the~$n$-th row and identically the~$n$-th column
of~$\m{C_{M}}$ is the self-capacitance (to infinity)~$C_{nn}$
of the $n$-th node. An abstract conductor at infinity serves as the
effective datum of the schematic corresponding to the Maxwell matrix.
The matrix is expressed in the basis of node fluxes~$\m{\Phi}_{\infty}$
referenced to that infinity conductor, assumed at zero potential;
note, $\dim\m{\Phi}_{\infty}=\left|\mathcal{N}\right|\times1=\left(N+1\right)\times1$. 

By selecting~$n_{0}$ as the physical ground node and datum of the
circuit, we set its respective flux~$\Phi_{n_{0}}$ to zero. We need
to thus eliminate~$\Phi_{n_{0}}$ as a degree of freedom and reduce
the flux basis to an~$N$-length column vector of node-to-datum fluxes~$\m{\Phi_{n}}$;
note, $\dim\m{\Phi_{n}}=N\times1$. We perform the basis reduction~$\m{\Phi}_{\infty}=\m S_{N}\m{\Phi_{n}}$
with the nearly-trivial linear transformation~$\m S_{N}=\left[\begin{array}{cc}
\m 0_{N} & \m I_{N}\end{array}\right]^{\intercal}$, where~$\m 0_{N}$ is the column vector of all zeros with length~$N$
and~$\m I_{N}$ is the identity matrix of dimension~$N$. In the
reduced basis~$\m{\Phi_{n}}$, the reduced capacitance matrix of
the cell is~
\[
\m C_{\mathrm{cell}}=\m S_{N}^{\intercal}\m{C_{M}}\m S_{N}\;.
\]
Due to the simple form of~$\m S_{N}$, this is equivalent to simply
dropping the first row and column of~$\m{C_{M}}$; i.e.,
\[
\m C_{\mathrm{cell}}=\block\left(\m{C_{M}},\mathcal{N}_{\mathrm{cell}},\mathcal{N}_{\mathrm{cell}}\right)\;,
\]
where the set of non-datum nodes of the cell is~$\mathcal{N}_{\mathrm{cell}}=\mathcal{N}-\left\{ n_{0}\right\} $.

\bibliographystyle{apsrev4-1}
\bibliography{library}

\begin{thebibliography}{65}%
\makeatletter
\providecommand \@ifxundefined [1]{%
 \@ifx{#1\undefined}
}%
\providecommand \@ifnum [1]{%
 \ifnum #1\expandafter \@firstoftwo
 \else \expandafter \@secondoftwo
 \fi
}%
\providecommand \@ifx [1]{%
 \ifx #1\expandafter \@firstoftwo
 \else \expandafter \@secondoftwo
 \fi
}%
\providecommand \natexlab [1]{#1}%
\providecommand \enquote  [1]{``#1''}%
\providecommand \bibnamefont  [1]{#1}%
\providecommand \bibfnamefont [1]{#1}%
\providecommand \citenamefont [1]{#1}%
\providecommand \href@noop [0]{\@secondoftwo}%
\providecommand \href [0]{\begingroup \@sanitize@url \@href}%
\providecommand \@href[1]{\@@startlink{#1}\@@href}%
\providecommand \@@href[1]{\endgroup#1\@@endlink}%
\providecommand \@sanitize@url [0]{\catcode `\\12\catcode `\$12\catcode
  `\&12\catcode `\#12\catcode `\^12\catcode `\_12\catcode `\%12\relax}%
\providecommand \@@startlink[1]{}%
\providecommand \@@endlink[0]{}%
\providecommand \url  [0]{\begingroup\@sanitize@url \@url }%
\providecommand \@url [1]{\endgroup\@href {#1}{\urlprefix }}%
\providecommand \urlprefix  [0]{URL }%
\providecommand \Eprint [0]{\href }%
\providecommand \doibase [0]{http://dx.doi.org/}%
\providecommand \selectlanguage [0]{\@gobble}%
\providecommand \bibinfo  [0]{\@secondoftwo}%
\providecommand \bibfield  [0]{\@secondoftwo}%
\providecommand \translation [1]{[#1]}%
\providecommand \BibitemOpen [0]{}%
\providecommand \bibitemStop [0]{}%
\providecommand \bibitemNoStop [0]{.\EOS\space}%
\providecommand \EOS [0]{\spacefactor3000\relax}%
\providecommand \BibitemShut  [1]{\csname bibitem#1\endcsname}%
\let\auto@bib@innerbib\@empty
\bibitem [{\citenamefont {Devoret}\ and\ \citenamefont
  {Schoelkopf}(2013)}]{Devoret2013}%
  \BibitemOpen
  \bibfield  {author} {\bibinfo {author} {\bibfnamefont {M.~H.}\ \bibnamefont
  {Devoret}}\ and\ \bibinfo {author} {\bibfnamefont {R.~J.}\ \bibnamefont
  {Schoelkopf}},\ }\href {\doibase 10.1126/science.1231930} {\bibfield
  {journal} {\bibinfo  {journal} {Science}\ }\textbf {\bibinfo {volume}
  {339}},\ \bibinfo {pages} {1169} (\bibinfo {year} {2013})}\BibitemShut
  {NoStop}%
\bibitem [{\citenamefont {Gambetta}\ \emph {et~al.}(2017)\citenamefont
  {Gambetta}, \citenamefont {Chow},\ and\ \citenamefont
  {Steffen}}]{Gambetta2017}%
  \BibitemOpen
  \bibfield  {author} {\bibinfo {author} {\bibfnamefont {J.~M.}\ \bibnamefont
  {Gambetta}}, \bibinfo {author} {\bibfnamefont {J.~M.}\ \bibnamefont {Chow}},
  \ and\ \bibinfo {author} {\bibfnamefont {M.}~\bibnamefont {Steffen}},\ }\href
  {\doibase 10.1038/s41534-016-0004-0} {\bibfield  {journal} {\bibinfo
  {journal} {npj Quantum Information}\ }\textbf {\bibinfo {volume} {3}},\
  \bibinfo {pages} {2} (\bibinfo {year} {2017})}\BibitemShut {NoStop}%
\bibitem [{\citenamefont {Krantz}\ \emph {et~al.}(2019)\citenamefont {Krantz},
  \citenamefont {Kjaergaard}, \citenamefont {Yan}, \citenamefont {Orlando},
  \citenamefont {Gustavsson},\ and\ \citenamefont {Oliver}}]{Krantz2019}%
  \BibitemOpen
  \bibfield  {author} {\bibinfo {author} {\bibfnamefont {P.}~\bibnamefont
  {Krantz}}, \bibinfo {author} {\bibfnamefont {M.}~\bibnamefont {Kjaergaard}},
  \bibinfo {author} {\bibfnamefont {F.}~\bibnamefont {Yan}}, \bibinfo {author}
  {\bibfnamefont {T.~P.}\ \bibnamefont {Orlando}}, \bibinfo {author}
  {\bibfnamefont {S.}~\bibnamefont {Gustavsson}}, \ and\ \bibinfo {author}
  {\bibfnamefont {W.~D.}\ \bibnamefont {Oliver}},\ }\href {\doibase
  10.1063/1.5089550} {\bibfield  {journal} {\bibinfo  {journal} {Applied
  Physics Reviews}\ }\textbf {\bibinfo {volume} {6}},\ \bibinfo {pages}
  {021318} (\bibinfo {year} {2019})}\BibitemShut {NoStop}%
\bibitem [{\citenamefont {Kjaergaard}\ \emph {et~al.}(2020)\citenamefont
  {Kjaergaard}, \citenamefont {Schwartz}, \citenamefont {Braum{\"{u}}ller},
  \citenamefont {Krantz}, \citenamefont {Wang}, \citenamefont {Gustavsson},\
  and\ \citenamefont {Oliver}}]{Kjaergaard2020}%
  \BibitemOpen
  \bibfield  {author} {\bibinfo {author} {\bibfnamefont {M.}~\bibnamefont
  {Kjaergaard}}, \bibinfo {author} {\bibfnamefont {M.~E.}\ \bibnamefont
  {Schwartz}}, \bibinfo {author} {\bibfnamefont {J.}~\bibnamefont
  {Braum{\"{u}}ller}}, \bibinfo {author} {\bibfnamefont {P.}~\bibnamefont
  {Krantz}}, \bibinfo {author} {\bibfnamefont {J.~I.}\ \bibnamefont {Wang}},
  \bibinfo {author} {\bibfnamefont {S.}~\bibnamefont {Gustavsson}}, \ and\
  \bibinfo {author} {\bibfnamefont {W.~D.}\ \bibnamefont {Oliver}},\ }\href
  {\doibase 10.1146/annurev-conmatphys-031119-050605} {\enquote {\bibinfo
  {title} {{Superconducting Qubits: Current State of Play}},}\ } (\bibinfo
  {year} {2020}),\ \Eprint {http://arxiv.org/abs/1905.13641} {arXiv:1905.13641}
  \BibitemShut {NoStop}%
\bibitem [{\citenamefont {Blais}\ \emph {et~al.}(2020)\citenamefont {Blais},
  \citenamefont {Grimsmo}, \citenamefont {Girvin},\ and\ \citenamefont
  {Wallraff}}]{Blais2020}%
  \BibitemOpen
  \bibfield  {author} {\bibinfo {author} {\bibfnamefont {A.}~\bibnamefont
  {Blais}}, \bibinfo {author} {\bibfnamefont {A.~L.}\ \bibnamefont {Grimsmo}},
  \bibinfo {author} {\bibfnamefont {S.~M.}\ \bibnamefont {Girvin}}, \ and\
  \bibinfo {author} {\bibfnamefont {A.}~\bibnamefont {Wallraff}},\ }\href
  {http://arxiv.org/abs/2005.12667} {\  (\bibinfo {year} {2020})},\ \Eprint
  {http://arxiv.org/abs/2005.12667} {arXiv:2005.12667} \BibitemShut {NoStop}%
\bibitem [{\citenamefont {Nigg}\ \emph {et~al.}(2012)\citenamefont {Nigg},
  \citenamefont {Paik}, \citenamefont {Vlastakis}, \citenamefont {Kirchmair},
  \citenamefont {Shankar}, \citenamefont {Frunzio}, \citenamefont {Devoret},
  \citenamefont {Schoelkopf},\ and\ \citenamefont {Girvin}}]{Nigg2012}%
  \BibitemOpen
  \bibfield  {author} {\bibinfo {author} {\bibfnamefont {S.~E.}\ \bibnamefont
  {Nigg}}, \bibinfo {author} {\bibfnamefont {H.}~\bibnamefont {Paik}}, \bibinfo
  {author} {\bibfnamefont {B.}~\bibnamefont {Vlastakis}}, \bibinfo {author}
  {\bibfnamefont {G.}~\bibnamefont {Kirchmair}}, \bibinfo {author}
  {\bibfnamefont {S.}~\bibnamefont {Shankar}}, \bibinfo {author} {\bibfnamefont
  {L.}~\bibnamefont {Frunzio}}, \bibinfo {author} {\bibfnamefont {M.~H.}\
  \bibnamefont {Devoret}}, \bibinfo {author} {\bibfnamefont {R.~J.}\
  \bibnamefont {Schoelkopf}}, \ and\ \bibinfo {author} {\bibfnamefont {S.~M.}\
  \bibnamefont {Girvin}},\ }\href {\doibase 10.1103/PhysRevLett.108.240502}
  {\bibfield  {journal} {\bibinfo  {journal} {Physical Review Letters}\
  }\textbf {\bibinfo {volume} {108}},\ \bibinfo {pages} {240502} (\bibinfo
  {year} {2012})}\BibitemShut {NoStop}%
\bibitem [{\citenamefont {Bourassa}\ \emph {et~al.}(2012)\citenamefont
  {Bourassa}, \citenamefont {Beaudoin}, \citenamefont {Gambetta},\ and\
  \citenamefont {Blais}}]{Bourassa2012}%
  \BibitemOpen
  \bibfield  {author} {\bibinfo {author} {\bibfnamefont {J.}~\bibnamefont
  {Bourassa}}, \bibinfo {author} {\bibfnamefont {F.}~\bibnamefont {Beaudoin}},
  \bibinfo {author} {\bibfnamefont {J.~M.}\ \bibnamefont {Gambetta}}, \ and\
  \bibinfo {author} {\bibfnamefont {A.}~\bibnamefont {Blais}},\ }\href
  {\doibase 10.1103/PhysRevA.86.013814} {\bibfield  {journal} {\bibinfo
  {journal} {Physical Review A}\ }\textbf {\bibinfo {volume} {86}},\ \bibinfo
  {pages} {013814} (\bibinfo {year} {2012})},\ \Eprint
  {http://arxiv.org/abs/arXiv:1204.2237v2} {arXiv:arXiv:1204.2237v2}
  \BibitemShut {NoStop}%
\bibitem [{\citenamefont {Solgun}\ \emph {et~al.}(2014)\citenamefont {Solgun},
  \citenamefont {Abraham},\ and\ \citenamefont {DiVincenzo}}]{Solgun2014}%
  \BibitemOpen
  \bibfield  {author} {\bibinfo {author} {\bibfnamefont {F.}~\bibnamefont
  {Solgun}}, \bibinfo {author} {\bibfnamefont {D.~W.}\ \bibnamefont {Abraham}},
  \ and\ \bibinfo {author} {\bibfnamefont {D.~P.}\ \bibnamefont {DiVincenzo}},\
  }\href {\doibase 10.1103/PhysRevB.90.134504} {\bibfield  {journal} {\bibinfo
  {journal} {Physical Review B}\ }\textbf {\bibinfo {volume} {90}},\ \bibinfo
  {pages} {134504} (\bibinfo {year} {2014})}\BibitemShut {NoStop}%
\bibitem [{\citenamefont {Solgun}\ and\ \citenamefont
  {DiVincenzo}(2015)}]{Solgun2015}%
  \BibitemOpen
  \bibfield  {author} {\bibinfo {author} {\bibfnamefont {F.}~\bibnamefont
  {Solgun}}\ and\ \bibinfo {author} {\bibfnamefont {D.~P.}\ \bibnamefont
  {DiVincenzo}},\ }\href {\doibase 10.1016/j.aop.2015.07.005} {\bibfield
  {journal} {\bibinfo  {journal} {Annals of Physics}\ }\textbf {\bibinfo
  {volume} {361}},\ \bibinfo {pages} {605} (\bibinfo {year} {2015})},\ \Eprint
  {http://arxiv.org/abs/1505.04116} {arXiv:1505.04116} \BibitemShut {NoStop}%
\bibitem [{\citenamefont {Smith}\ \emph {et~al.}(2016)\citenamefont {Smith},
  \citenamefont {Kou}, \citenamefont {Vool}, \citenamefont {Pop}, \citenamefont
  {Frunzio}, \citenamefont {Schoelkopf},\ and\ \citenamefont
  {Devoret}}]{Smith2016}%
  \BibitemOpen
  \bibfield  {author} {\bibinfo {author} {\bibfnamefont {W.~C.}\ \bibnamefont
  {Smith}}, \bibinfo {author} {\bibfnamefont {A.}~\bibnamefont {Kou}}, \bibinfo
  {author} {\bibfnamefont {U.}~\bibnamefont {Vool}}, \bibinfo {author}
  {\bibfnamefont {I.~M.}\ \bibnamefont {Pop}}, \bibinfo {author} {\bibfnamefont
  {L.}~\bibnamefont {Frunzio}}, \bibinfo {author} {\bibfnamefont {R.~J.}\
  \bibnamefont {Schoelkopf}}, \ and\ \bibinfo {author} {\bibfnamefont {M.~H.}\
  \bibnamefont {Devoret}},\ }\href {\doibase 10.1103/PhysRevB.94.144507}
  {\bibfield  {journal} {\bibinfo  {journal} {Physical Review B}\ }\textbf
  {\bibinfo {volume} {94}},\ \bibinfo {pages} {144507} (\bibinfo {year}
  {2016})},\ \Eprint {http://arxiv.org/abs/1602.01793} {arXiv:1602.01793}
  \BibitemShut {NoStop}%
\bibitem [{\citenamefont {Malekakhlagh}\ and\ \citenamefont
  {T{\"{u}}reci}(2016)}]{Malekakhlagh2016-A2}%
  \BibitemOpen
  \bibfield  {author} {\bibinfo {author} {\bibfnamefont {M.}~\bibnamefont
  {Malekakhlagh}}\ and\ \bibinfo {author} {\bibfnamefont {H.~E.}\ \bibnamefont
  {T{\"{u}}reci}},\ }\href {\doibase 10.1103/PhysRevA.93.012120} {\bibfield
  {journal} {\bibinfo  {journal} {Physical Review A}\ }\textbf {\bibinfo
  {volume} {93}},\ \bibinfo {pages} {012120} (\bibinfo {year} {2016})},\
  \Eprint {http://arxiv.org/abs/1506.02773} {arXiv:1506.02773} \BibitemShut
  {NoStop}%
\bibitem [{\citenamefont {Gely}\ \emph {et~al.}(2017)\citenamefont {Gely},
  \citenamefont {Parra-Rodriguez}, \citenamefont {Bothner}, \citenamefont
  {Blanter}, \citenamefont {Bosman}, \citenamefont {Solano},\ and\
  \citenamefont {Steele}}]{Gely2017}%
  \BibitemOpen
  \bibfield  {author} {\bibinfo {author} {\bibfnamefont {M.~F.}\ \bibnamefont
  {Gely}}, \bibinfo {author} {\bibfnamefont {A.}~\bibnamefont
  {Parra-Rodriguez}}, \bibinfo {author} {\bibfnamefont {D.}~\bibnamefont
  {Bothner}}, \bibinfo {author} {\bibfnamefont {Y.~M.}\ \bibnamefont
  {Blanter}}, \bibinfo {author} {\bibfnamefont {S.~J.}\ \bibnamefont {Bosman}},
  \bibinfo {author} {\bibfnamefont {E.}~\bibnamefont {Solano}}, \ and\ \bibinfo
  {author} {\bibfnamefont {G.~A.}\ \bibnamefont {Steele}},\ }\href {\doibase
  10.1103/PhysRevB.95.245115} {\bibfield  {journal} {\bibinfo  {journal}
  {Physical Review B}\ }\textbf {\bibinfo {volume} {95}},\ \bibinfo {pages}
  {245115} (\bibinfo {year} {2017})}\BibitemShut {NoStop}%
\bibitem [{\citenamefont {Malekakhlagh}\ \emph {et~al.}(2017)\citenamefont
  {Malekakhlagh}, \citenamefont {Petrescu},\ and\ \citenamefont
  {T{\"{u}}reci}}]{Malekakhlagh2017-Cutoff-Free}%
  \BibitemOpen
  \bibfield  {author} {\bibinfo {author} {\bibfnamefont {M.}~\bibnamefont
  {Malekakhlagh}}, \bibinfo {author} {\bibfnamefont {A.}~\bibnamefont
  {Petrescu}}, \ and\ \bibinfo {author} {\bibfnamefont {H.~E.}\ \bibnamefont
  {T{\"{u}}reci}},\ }\href {\doibase 10.1103/PhysRevLett.119.073601} {\bibfield
   {journal} {\bibinfo  {journal} {Physical Review Letters}\ }\textbf {\bibinfo
  {volume} {119}},\ \bibinfo {pages} {073601} (\bibinfo {year}
  {2017})}\BibitemShut {NoStop}%
\bibitem [{\citenamefont {Pechal}\ and\ \citenamefont
  {Safavi-Naeini}(2017)}]{Pechal2017}%
  \BibitemOpen
  \bibfield  {author} {\bibinfo {author} {\bibfnamefont {M.}~\bibnamefont
  {Pechal}}\ and\ \bibinfo {author} {\bibfnamefont {A.~H.}\ \bibnamefont
  {Safavi-Naeini}},\ }\href {\doibase 10.1103/PhysRevA.96.042305} {\bibfield
  {journal} {\bibinfo  {journal} {Physical Review A}\ }\textbf {\bibinfo
  {volume} {96}},\ \bibinfo {pages} {042305} (\bibinfo {year} {2017})},\
  \Eprint {http://arxiv.org/abs/1706.05368} {arXiv:1706.05368} \BibitemShut
  {NoStop}%
\bibitem [{\citenamefont {Minev}\ \emph {et~al.}(2018)\citenamefont {Minev},
  \citenamefont {Leghtas}, \citenamefont {Mudhada}, \citenamefont {Reinhold},
  \citenamefont {Diringer},\ and\ \citenamefont {Devoret}}]{pyEPR}%
  \BibitemOpen
  \bibfield  {author} {\bibinfo {author} {\bibfnamefont {Z.~K.}\ \bibnamefont
  {Minev}}, \bibinfo {author} {\bibfnamefont {Z.}~\bibnamefont {Leghtas}},
  \bibinfo {author} {\bibfnamefont {S.~O.}\ \bibnamefont {Mudhada}}, \bibinfo
  {author} {\bibfnamefont {P.}~\bibnamefont {Reinhold}}, \bibinfo {author}
  {\bibfnamefont {A.}~\bibnamefont {Diringer}}, \ and\ \bibinfo {author}
  {\bibfnamefont {M.~H.}\ \bibnamefont {Devoret}},\ }\href@noop {} {\enquote
  {\bibinfo {title} {{pyEPR: The energy-participation-ratio (EPR) open-source
  framework for quantum device design}},}\ } (\bibinfo {year}
  {2018})\BibitemShut {NoStop}%
\bibitem [{\citenamefont {Parra-Rodriguez}\ \emph {et~al.}(2019)\citenamefont
  {Parra-Rodriguez}, \citenamefont {Egusquiza}, \citenamefont {DiVincenzo},\
  and\ \citenamefont {Solano}}]{Parra-Rodriguez2018}%
  \BibitemOpen
  \bibfield  {author} {\bibinfo {author} {\bibfnamefont {A.}~\bibnamefont
  {Parra-Rodriguez}}, \bibinfo {author} {\bibfnamefont {I.~L.}\ \bibnamefont
  {Egusquiza}}, \bibinfo {author} {\bibfnamefont {D.~P.}\ \bibnamefont
  {DiVincenzo}}, \ and\ \bibinfo {author} {\bibfnamefont {E.}~\bibnamefont
  {Solano}},\ }\href {\doibase 10.1103/PhysRevB.99.014514} {\bibfield
  {journal} {\bibinfo  {journal} {Physical Review B}\ }\textbf {\bibinfo
  {volume} {99}},\ \bibinfo {pages} {014514} (\bibinfo {year} {2019})},\
  \Eprint {http://arxiv.org/abs/1810.08471} {arXiv:1810.08471} \BibitemShut
  {NoStop}%
\bibitem [{\citenamefont {Parra-Rodriguez}\ \emph {et~al.}(2018)\citenamefont
  {Parra-Rodriguez}, \citenamefont {Rico}, \citenamefont {Solano},\ and\
  \citenamefont {Egusquiza}}]{Parra-Rodriguez2018a}%
  \BibitemOpen
  \bibfield  {author} {\bibinfo {author} {\bibfnamefont {A.}~\bibnamefont
  {Parra-Rodriguez}}, \bibinfo {author} {\bibfnamefont {E.}~\bibnamefont
  {Rico}}, \bibinfo {author} {\bibfnamefont {E.}~\bibnamefont {Solano}}, \ and\
  \bibinfo {author} {\bibfnamefont {I.~L.}\ \bibnamefont {Egusquiza}},\ }\href
  {\doibase 10.1088/2058-9565/aab1ba} {\bibfield  {journal} {\bibinfo
  {journal} {Quantum Science and Technology}\ }\textbf {\bibinfo {volume}
  {3}},\ \bibinfo {pages} {024012} (\bibinfo {year} {2018})},\ \Eprint
  {http://arxiv.org/abs/1711.08817} {arXiv:1711.08817} \BibitemShut {NoStop}%
\bibitem [{\citenamefont {Ansari}(2019)}]{Ansari2018}%
  \BibitemOpen
  \bibfield  {author} {\bibinfo {author} {\bibfnamefont {M.~H.}\ \bibnamefont
  {Ansari}},\ }\href {\doibase 10.1103/PhysRevB.100.024509} {\bibfield
  {journal} {\bibinfo  {journal} {Physical Review B}\ }\textbf {\bibinfo
  {volume} {100}},\ \bibinfo {pages} {024509} (\bibinfo {year} {2019})},\
  \Eprint {http://arxiv.org/abs/1807.00792} {arXiv:1807.00792} \BibitemShut
  {NoStop}%
\bibitem [{\citenamefont {Krupko}\ \emph {et~al.}(2018)\citenamefont {Krupko},
  \citenamefont {Nguyen}, \citenamefont {Wei{\ss}l}, \citenamefont {Dumur},
  \citenamefont {Puertas}, \citenamefont {Dassonneville}, \citenamefont {Naud},
  \citenamefont {Hekking}, \citenamefont {Basko}, \citenamefont {Buisson},
  \citenamefont {Roch},\ and\ \citenamefont {Hasch-Guichard}}]{Krupko2018}%
  \BibitemOpen
  \bibfield  {author} {\bibinfo {author} {\bibfnamefont {Y.}~\bibnamefont
  {Krupko}}, \bibinfo {author} {\bibfnamefont {V.~D.}\ \bibnamefont {Nguyen}},
  \bibinfo {author} {\bibfnamefont {T.}~\bibnamefont {Wei{\ss}l}}, \bibinfo
  {author} {\bibfnamefont {{\'{E}}.}~\bibnamefont {Dumur}}, \bibinfo {author}
  {\bibfnamefont {J.}~\bibnamefont {Puertas}}, \bibinfo {author} {\bibfnamefont
  {R.}~\bibnamefont {Dassonneville}}, \bibinfo {author} {\bibfnamefont
  {C.}~\bibnamefont {Naud}}, \bibinfo {author} {\bibfnamefont {F.~W.~J.}\
  \bibnamefont {Hekking}}, \bibinfo {author} {\bibfnamefont {D.~M.}\
  \bibnamefont {Basko}}, \bibinfo {author} {\bibfnamefont {O.}~\bibnamefont
  {Buisson}}, \bibinfo {author} {\bibfnamefont {N.}~\bibnamefont {Roch}}, \
  and\ \bibinfo {author} {\bibfnamefont {W.}~\bibnamefont {Hasch-Guichard}},\
  }\href {\doibase 10.1103/PhysRevB.98.094516} {\bibfield  {journal} {\bibinfo
  {journal} {Physical Review B}\ }\textbf {\bibinfo {volume} {98}},\ \bibinfo
  {pages} {094516} (\bibinfo {year} {2018})}\BibitemShut {NoStop}%
\bibitem [{\citenamefont {Malekakhlagh}\ \emph
  {et~al.}(2020{\natexlab{a}})\citenamefont {Malekakhlagh}, \citenamefont
  {Petrescu},\ and\ \citenamefont {T{\"{u}}reci}}]{Malekakhlagh2018}%
  \BibitemOpen
  \bibfield  {author} {\bibinfo {author} {\bibfnamefont {M.}~\bibnamefont
  {Malekakhlagh}}, \bibinfo {author} {\bibfnamefont {A.}~\bibnamefont
  {Petrescu}}, \ and\ \bibinfo {author} {\bibfnamefont {H.~E.}\ \bibnamefont
  {T{\"{u}}reci}},\ }\href {\doibase 10.1103/PhysRevB.101.134509} {\bibfield
  {journal} {\bibinfo  {journal} {Physical Review B}\ }\textbf {\bibinfo
  {volume} {101}},\ \bibinfo {pages} {134509} (\bibinfo {year}
  {2020}{\natexlab{a}})},\ \Eprint {http://arxiv.org/abs/1809.04667}
  {arXiv:1809.04667} \BibitemShut {NoStop}%
\bibitem [{\citenamefont {Solgun}\ \emph {et~al.}(2019)\citenamefont {Solgun},
  \citenamefont {DiVincenzo},\ and\ \citenamefont {Gambetta}}]{Solgun2017}%
  \BibitemOpen
  \bibfield  {author} {\bibinfo {author} {\bibfnamefont {F.}~\bibnamefont
  {Solgun}}, \bibinfo {author} {\bibfnamefont {D.~P.}\ \bibnamefont
  {DiVincenzo}}, \ and\ \bibinfo {author} {\bibfnamefont {J.~M.}\ \bibnamefont
  {Gambetta}},\ }\href {\doibase 10.1109/TMTT.2019.2893639} {\bibfield
  {journal} {\bibinfo  {journal} {IEEE Transactions on Microwave Theory and
  Techniques}\ }\textbf {\bibinfo {volume} {67}},\ \bibinfo {pages} {928}
  (\bibinfo {year} {2019})},\ \Eprint {http://arxiv.org/abs/1712.08154}
  {arXiv:1712.08154} \BibitemShut {NoStop}%
\bibitem [{\citenamefont {Petrescu}\ \emph {et~al.}(2019)\citenamefont
  {Petrescu}, \citenamefont {Malekakhlagh},\ and\ \citenamefont
  {T{\"{u}}reci}}]{Petrescu2019}%
  \BibitemOpen
  \bibfield  {author} {\bibinfo {author} {\bibfnamefont {A.}~\bibnamefont
  {Petrescu}}, \bibinfo {author} {\bibfnamefont {M.}~\bibnamefont
  {Malekakhlagh}}, \ and\ \bibinfo {author} {\bibfnamefont {H.~E.}\
  \bibnamefont {T{\"{u}}reci}},\ }\href {http://arxiv.org/abs/1908.01240} {\
  (\bibinfo {year} {2019})},\ \Eprint {http://arxiv.org/abs/1908.01240}
  {arXiv:1908.01240} \BibitemShut {NoStop}%
\bibitem [{\citenamefont {You}\ \emph {et~al.}(2019)\citenamefont {You},
  \citenamefont {Sauls},\ and\ \citenamefont {Koch}}]{You2019-Koch}%
  \BibitemOpen
  \bibfield  {author} {\bibinfo {author} {\bibfnamefont {X.}~\bibnamefont
  {You}}, \bibinfo {author} {\bibfnamefont {J.~A.}\ \bibnamefont {Sauls}}, \
  and\ \bibinfo {author} {\bibfnamefont {J.}~\bibnamefont {Koch}},\ }\href
  {\doibase 10.1103/PhysRevB.99.174512} {\bibfield  {journal} {\bibinfo
  {journal} {Physical Review B}\ }\textbf {\bibinfo {volume} {99}},\ \bibinfo
  {pages} {174512} (\bibinfo {year} {2019})},\ \Eprint
  {http://arxiv.org/abs/1902.04734} {arXiv:1902.04734} \BibitemShut {NoStop}%
\bibitem [{\citenamefont {{Di Paolo}}\ \emph {et~al.}(2019)\citenamefont {{Di
  Paolo}}, \citenamefont {Baker}, \citenamefont {Foley}, \citenamefont
  {S{\'{e}}n{\'{e}}chal},\ and\ \citenamefont {Blais}}]{DiPaolo2019}%
  \BibitemOpen
  \bibfield  {author} {\bibinfo {author} {\bibfnamefont {A.}~\bibnamefont {{Di
  Paolo}}}, \bibinfo {author} {\bibfnamefont {T.~E.}\ \bibnamefont {Baker}},
  \bibinfo {author} {\bibfnamefont {A.}~\bibnamefont {Foley}}, \bibinfo
  {author} {\bibfnamefont {D.}~\bibnamefont {S{\'{e}}n{\'{e}}chal}}, \ and\
  \bibinfo {author} {\bibfnamefont {A.}~\bibnamefont {Blais}},\ }\href
  {http://arxiv.org/abs/1912.01018} {\  (\bibinfo {year} {2019})},\ \Eprint
  {http://arxiv.org/abs/1912.01018} {arXiv:1912.01018} \BibitemShut {NoStop}%
\bibitem [{\citenamefont {Gely}\ and\ \citenamefont {Steele}(2020)}]{Gely2020}%
  \BibitemOpen
  \bibfield  {author} {\bibinfo {author} {\bibfnamefont {M.~F.}\ \bibnamefont
  {Gely}}\ and\ \bibinfo {author} {\bibfnamefont {G.~A.}\ \bibnamefont
  {Steele}},\ }\href {\doibase 10.1088/1367-2630/ab60f6} {\bibfield  {journal}
  {\bibinfo  {journal} {New Journal of Physics}\ }\textbf {\bibinfo {volume}
  {22}},\ \bibinfo {pages} {013025} (\bibinfo {year} {2020})}\BibitemShut
  {NoStop}%
\bibitem [{\citenamefont {Ding}\ \emph {et~al.}(2020)\citenamefont {Ding},
  \citenamefont {Ku}, \citenamefont {Shi},\ and\ \citenamefont
  {Zhao}}]{Ding2020}%
  \BibitemOpen
  \bibfield  {author} {\bibinfo {author} {\bibfnamefont {D.}~\bibnamefont
  {Ding}}, \bibinfo {author} {\bibfnamefont {H.~S.}\ \bibnamefont {Ku}},
  \bibinfo {author} {\bibfnamefont {Y.}~\bibnamefont {Shi}}, \ and\ \bibinfo
  {author} {\bibfnamefont {H.~H.}\ \bibnamefont {Zhao}},\ }\href@noop {}
  {\enquote {\bibinfo {title} {{Free mode removal and mode decoupling for
  simulating general superconducting quantum circuits}},}\ } (\bibinfo {year}
  {2020}),\ \Eprint {http://arxiv.org/abs/2011.10564} {arXiv:2011.10564}
  \BibitemShut {NoStop}%
\bibitem [{\citenamefont {Kerman}(2020)}]{Kerman2020}%
  \BibitemOpen
  \bibfield  {author} {\bibinfo {author} {\bibfnamefont {A.~J.}\ \bibnamefont
  {Kerman}},\ }\href {https://arxiv.org/abs/2010.14929
  http://arxiv.org/abs/2010.14929} {\  (\bibinfo {year} {2020})},\ \Eprint
  {http://arxiv.org/abs/2010.14929} {arXiv:2010.14929} \BibitemShut {NoStop}%
\bibitem [{\citenamefont {Minev}\ \emph {et~al.}(2020)\citenamefont {Minev},
  \citenamefont {Leghtas}, \citenamefont {Mundhada}, \citenamefont
  {Christakis}, \citenamefont {Pop},\ and\ \citenamefont
  {Devoret}}]{Minev2020}%
  \BibitemOpen
  \bibfield  {author} {\bibinfo {author} {\bibfnamefont {Z.~K.}\ \bibnamefont
  {Minev}}, \bibinfo {author} {\bibfnamefont {Z.}~\bibnamefont {Leghtas}},
  \bibinfo {author} {\bibfnamefont {S.~O.}\ \bibnamefont {Mundhada}}, \bibinfo
  {author} {\bibfnamefont {L.}~\bibnamefont {Christakis}}, \bibinfo {author}
  {\bibfnamefont {I.~M.}\ \bibnamefont {Pop}}, \ and\ \bibinfo {author}
  {\bibfnamefont {M.~H.}\ \bibnamefont {Devoret}},\ }\href
  {http://arxiv.org/abs/2010.00620} {\bibfield  {journal} {\bibinfo  {journal}
  {arXiv}\ } (\bibinfo {year} {2020})},\ \Eprint
  {http://arxiv.org/abs/2010.00620} {arXiv:2010.00620} \BibitemShut {NoStop}%
\bibitem [{\citenamefont {Kyaw}\ \emph {et~al.}(2020)\citenamefont {Kyaw},
  \citenamefont {Menke}, \citenamefont {Sim}, \citenamefont {Sawaya},
  \citenamefont {Oliver}, \citenamefont {Guerreschi},\ and\ \citenamefont
  {Aspuru-Guzik}}]{Kyaw2020}%
  \BibitemOpen
  \bibfield  {author} {\bibinfo {author} {\bibfnamefont {T.~H.}\ \bibnamefont
  {Kyaw}}, \bibinfo {author} {\bibfnamefont {T.}~\bibnamefont {Menke}},
  \bibinfo {author} {\bibfnamefont {S.}~\bibnamefont {Sim}}, \bibinfo {author}
  {\bibfnamefont {N.~P.~D.}\ \bibnamefont {Sawaya}}, \bibinfo {author}
  {\bibfnamefont {W.~D.}\ \bibnamefont {Oliver}}, \bibinfo {author}
  {\bibfnamefont {G.~G.}\ \bibnamefont {Guerreschi}}, \ and\ \bibinfo {author}
  {\bibfnamefont {A.}~\bibnamefont {Aspuru-Guzik}},\ }\href
  {http://arxiv.org/abs/2006.03070} {\  (\bibinfo {year} {2020})},\ \Eprint
  {http://arxiv.org/abs/2006.03070} {arXiv:2006.03070} \BibitemShut {NoStop}%
\bibitem [{\citenamefont {Malekakhlagh}\ \emph
  {et~al.}(2020{\natexlab{b}})\citenamefont {Malekakhlagh}, \citenamefont
  {Magesan},\ and\ \citenamefont {McKay}}]{Malekakhlagh2020}%
  \BibitemOpen
  \bibfield  {author} {\bibinfo {author} {\bibfnamefont {M.}~\bibnamefont
  {Malekakhlagh}}, \bibinfo {author} {\bibfnamefont {E.}~\bibnamefont
  {Magesan}}, \ and\ \bibinfo {author} {\bibfnamefont {D.~C.}\ \bibnamefont
  {McKay}},\ }\href {http://arxiv.org/abs/2005.00133} {\  (\bibinfo {year}
  {2020}{\natexlab{b}})},\ \Eprint {http://arxiv.org/abs/2005.00133}
  {arXiv:2005.00133} \BibitemShut {NoStop}%
\bibitem [{\citenamefont {Menke}\ \emph {et~al.}(2021)\citenamefont {Menke},
  \citenamefont {H{\"{a}}se}, \citenamefont {Gustavsson}, \citenamefont
  {Kerman}, \citenamefont {Oliver},\ and\ \citenamefont
  {Aspuru-Guzik}}]{Menke2021}%
  \BibitemOpen
  \bibfield  {author} {\bibinfo {author} {\bibfnamefont {T.}~\bibnamefont
  {Menke}}, \bibinfo {author} {\bibfnamefont {F.}~\bibnamefont {H{\"{a}}se}},
  \bibinfo {author} {\bibfnamefont {S.}~\bibnamefont {Gustavsson}}, \bibinfo
  {author} {\bibfnamefont {A.~J.}\ \bibnamefont {Kerman}}, \bibinfo {author}
  {\bibfnamefont {W.~D.}\ \bibnamefont {Oliver}}, \ and\ \bibinfo {author}
  {\bibfnamefont {A.}~\bibnamefont {Aspuru-Guzik}},\ }\href {\doibase
  10.1038/s41534-021-00382-6} {\bibfield  {journal} {\bibinfo  {journal} {npj
  Quantum Information}\ }\textbf {\bibinfo {volume} {7}},\ \bibinfo {pages}
  {49} (\bibinfo {year} {2021})},\ \Eprint {http://arxiv.org/abs/1912.03322}
  {arXiv:1912.03322} \BibitemShut {NoStop}%
\bibitem [{\citenamefont {Wallraff}\ \emph {et~al.}(2004)\citenamefont
  {Wallraff}, \citenamefont {Schuster}, \citenamefont {Blais}, \citenamefont
  {Frunzio}, \citenamefont {Huang}, \citenamefont {Majer}, \citenamefont
  {Kumar}, \citenamefont {Girvin},\ and\ \citenamefont
  {Schoelkopf}}]{Wallraff2004}%
  \BibitemOpen
  \bibfield  {author} {\bibinfo {author} {\bibfnamefont {A.}~\bibnamefont
  {Wallraff}}, \bibinfo {author} {\bibfnamefont {D.~I.}\ \bibnamefont
  {Schuster}}, \bibinfo {author} {\bibfnamefont {A.}~\bibnamefont {Blais}},
  \bibinfo {author} {\bibfnamefont {L.}~\bibnamefont {Frunzio}}, \bibinfo
  {author} {\bibfnamefont {R.-S.}\ \bibnamefont {Huang}}, \bibinfo {author}
  {\bibfnamefont {J.}~\bibnamefont {Majer}}, \bibinfo {author} {\bibfnamefont
  {S.}~\bibnamefont {Kumar}}, \bibinfo {author} {\bibfnamefont {S.~M.}\
  \bibnamefont {Girvin}}, \ and\ \bibinfo {author} {\bibfnamefont {R.~J.}\
  \bibnamefont {Schoelkopf}},\ }\href {\doibase 10.1038/nature02851} {\bibfield
   {journal} {\bibinfo  {journal} {Nature}\ }\textbf {\bibinfo {volume}
  {431}},\ \bibinfo {pages} {162} (\bibinfo {year} {2004})},\ \Eprint
  {http://arxiv.org/abs/0407325} {arXiv:0407325 [cond-mat]} \BibitemShut
  {NoStop}%
\bibitem [{\citenamefont {Paik}\ \emph {et~al.}(2011)\citenamefont {Paik},
  \citenamefont {Schuster}, \citenamefont {Bishop}, \citenamefont {Kirchmair},
  \citenamefont {Catelani}, \citenamefont {Sears}, \citenamefont {Johnson},
  \citenamefont {Reagor}, \citenamefont {Frunzio}, \citenamefont {Glazman},
  \citenamefont {Girvin}, \citenamefont {Devoret},\ and\ \citenamefont
  {Schoelkopf}}]{Paik2011}%
  \BibitemOpen
  \bibfield  {author} {\bibinfo {author} {\bibfnamefont {H.}~\bibnamefont
  {Paik}}, \bibinfo {author} {\bibfnamefont {D.~I.}\ \bibnamefont {Schuster}},
  \bibinfo {author} {\bibfnamefont {L.~S.}\ \bibnamefont {Bishop}}, \bibinfo
  {author} {\bibfnamefont {G.}~\bibnamefont {Kirchmair}}, \bibinfo {author}
  {\bibfnamefont {G.}~\bibnamefont {Catelani}}, \bibinfo {author}
  {\bibfnamefont {A.~P.}\ \bibnamefont {Sears}}, \bibinfo {author}
  {\bibfnamefont {B.~R.}\ \bibnamefont {Johnson}}, \bibinfo {author}
  {\bibfnamefont {M.~J.}\ \bibnamefont {Reagor}}, \bibinfo {author}
  {\bibfnamefont {L.}~\bibnamefont {Frunzio}}, \bibinfo {author} {\bibfnamefont
  {L.~I.}\ \bibnamefont {Glazman}}, \bibinfo {author} {\bibfnamefont {S.~M.}\
  \bibnamefont {Girvin}}, \bibinfo {author} {\bibfnamefont {M.~H.}\
  \bibnamefont {Devoret}}, \ and\ \bibinfo {author} {\bibfnamefont {R.~J.}\
  \bibnamefont {Schoelkopf}},\ }\href {\doibase 10.1103/PhysRevLett.107.240501}
  {\bibfield  {journal} {\bibinfo  {journal} {Physical Review Letters}\
  }\textbf {\bibinfo {volume} {107}},\ \bibinfo {pages} {240501} (\bibinfo
  {year} {2011})},\ \Eprint {http://arxiv.org/abs/1105.4652} {arXiv:1105.4652}
  \BibitemShut {NoStop}%
\bibitem [{\citenamefont {Barends}\ \emph {et~al.}(2013)\citenamefont
  {Barends}, \citenamefont {Kelly}, \citenamefont {Megrant}, \citenamefont
  {Sank}, \citenamefont {Jeffrey}, \citenamefont {Chen}, \citenamefont {Yin},
  \citenamefont {Chiaro}, \citenamefont {Mutus}, \citenamefont {Neill},
  \citenamefont {O'Malley}, \citenamefont {Roushan}, \citenamefont {Wenner},
  \citenamefont {White}, \citenamefont {Cleland},\ and\ \citenamefont
  {Martinis}}]{Barends2013}%
  \BibitemOpen
  \bibfield  {author} {\bibinfo {author} {\bibfnamefont {R.}~\bibnamefont
  {Barends}}, \bibinfo {author} {\bibfnamefont {J.}~\bibnamefont {Kelly}},
  \bibinfo {author} {\bibfnamefont {A.}~\bibnamefont {Megrant}}, \bibinfo
  {author} {\bibfnamefont {D.}~\bibnamefont {Sank}}, \bibinfo {author}
  {\bibfnamefont {E.}~\bibnamefont {Jeffrey}}, \bibinfo {author} {\bibfnamefont
  {Y.}~\bibnamefont {Chen}}, \bibinfo {author} {\bibfnamefont {Y.}~\bibnamefont
  {Yin}}, \bibinfo {author} {\bibfnamefont {B.}~\bibnamefont {Chiaro}},
  \bibinfo {author} {\bibfnamefont {J.}~\bibnamefont {Mutus}}, \bibinfo
  {author} {\bibfnamefont {C.}~\bibnamefont {Neill}}, \bibinfo {author}
  {\bibfnamefont {P.}~\bibnamefont {O'Malley}}, \bibinfo {author}
  {\bibfnamefont {P.}~\bibnamefont {Roushan}}, \bibinfo {author} {\bibfnamefont
  {J.}~\bibnamefont {Wenner}}, \bibinfo {author} {\bibfnamefont {T.~C.}\
  \bibnamefont {White}}, \bibinfo {author} {\bibfnamefont {A.~N.}\ \bibnamefont
  {Cleland}}, \ and\ \bibinfo {author} {\bibfnamefont {J.~M.}\ \bibnamefont
  {Martinis}},\ }\href {\doibase 10.1103/PhysRevLett.111.080502} {\bibfield
  {journal} {\bibinfo  {journal} {Physical Review Letters}\ }\textbf {\bibinfo
  {volume} {111}},\ \bibinfo {pages} {080502} (\bibinfo {year} {2013})},\
  \Eprint {http://arxiv.org/abs/1304.2322} {arXiv:1304.2322} \BibitemShut
  {NoStop}%
\bibitem [{\citenamefont {Minev}\ \emph {et~al.}(2013)\citenamefont {Minev},
  \citenamefont {Pop},\ and\ \citenamefont {Devoret}}]{Minev2013}%
  \BibitemOpen
  \bibfield  {author} {\bibinfo {author} {\bibfnamefont {Z.}~\bibnamefont
  {Minev}}, \bibinfo {author} {\bibfnamefont {I.~M.}\ \bibnamefont {Pop}}, \
  and\ \bibinfo {author} {\bibfnamefont {M.~H.}\ \bibnamefont {Devoret}},\
  }\href {\doibase 10.1063/1.4824201} {\bibfield  {journal} {\bibinfo
  {journal} {Applied Physics Letters}\ }\textbf {\bibinfo {volume} {103}},\
  \bibinfo {pages} {142604} (\bibinfo {year} {2013})},\ \Eprint
  {http://arxiv.org/abs/arXiv:1308.1743v1} {arXiv:arXiv:1308.1743v1}
  \BibitemShut {NoStop}%
\bibitem [{\citenamefont {Yan}\ \emph {et~al.}(2016)\citenamefont {Yan},
  \citenamefont {Gustavsson}, \citenamefont {Kamal}, \citenamefont {Birenbaum},
  \citenamefont {Sears}, \citenamefont {Hover}, \citenamefont {Gudmundsen},
  \citenamefont {Rosenberg}, \citenamefont {Samach}, \citenamefont {Weber},
  \citenamefont {Yoder}, \citenamefont {Orlando}, \citenamefont {Clarke},
  \citenamefont {Kerman},\ and\ \citenamefont {Oliver}}]{FYan2016}%
  \BibitemOpen
  \bibfield  {author} {\bibinfo {author} {\bibfnamefont {F.}~\bibnamefont
  {Yan}}, \bibinfo {author} {\bibfnamefont {S.}~\bibnamefont {Gustavsson}},
  \bibinfo {author} {\bibfnamefont {A.}~\bibnamefont {Kamal}}, \bibinfo
  {author} {\bibfnamefont {J.}~\bibnamefont {Birenbaum}}, \bibinfo {author}
  {\bibfnamefont {A.~P.}\ \bibnamefont {Sears}}, \bibinfo {author}
  {\bibfnamefont {D.}~\bibnamefont {Hover}}, \bibinfo {author} {\bibfnamefont
  {T.~J.}\ \bibnamefont {Gudmundsen}}, \bibinfo {author} {\bibfnamefont
  {D.}~\bibnamefont {Rosenberg}}, \bibinfo {author} {\bibfnamefont
  {G.}~\bibnamefont {Samach}}, \bibinfo {author} {\bibfnamefont
  {S.}~\bibnamefont {Weber}}, \bibinfo {author} {\bibfnamefont {J.~L.}\
  \bibnamefont {Yoder}}, \bibinfo {author} {\bibfnamefont {T.~P.}\ \bibnamefont
  {Orlando}}, \bibinfo {author} {\bibfnamefont {J.}~\bibnamefont {Clarke}},
  \bibinfo {author} {\bibfnamefont {A.~J.}\ \bibnamefont {Kerman}}, \ and\
  \bibinfo {author} {\bibfnamefont {W.~D.}\ \bibnamefont {Oliver}},\ }\href
  {\doibase 10.1038/ncomms12964} {\bibfield  {journal} {\bibinfo  {journal}
  {Nature Communications}\ }\textbf {\bibinfo {volume} {7}},\ \bibinfo {pages}
  {12964} (\bibinfo {year} {2016})}\BibitemShut {NoStop}%
\bibitem [{\citenamefont {Brecht}\ \emph {et~al.}(2016)\citenamefont {Brecht},
  \citenamefont {Pfaff}, \citenamefont {Wang}, \citenamefont {Chu},
  \citenamefont {Frunzio}, \citenamefont {Devoret},\ and\ \citenamefont
  {Schoelkopf}}]{Brecht2016}%
  \BibitemOpen
  \bibfield  {author} {\bibinfo {author} {\bibfnamefont {T.}~\bibnamefont
  {Brecht}}, \bibinfo {author} {\bibfnamefont {W.}~\bibnamefont {Pfaff}},
  \bibinfo {author} {\bibfnamefont {C.}~\bibnamefont {Wang}}, \bibinfo {author}
  {\bibfnamefont {Y.}~\bibnamefont {Chu}}, \bibinfo {author} {\bibfnamefont
  {L.}~\bibnamefont {Frunzio}}, \bibinfo {author} {\bibfnamefont {M.~H.}\
  \bibnamefont {Devoret}}, \ and\ \bibinfo {author} {\bibfnamefont {R.~J.}\
  \bibnamefont {Schoelkopf}},\ }\href {\doibase 10.1038/npjqi.2016.2}
  {\bibfield  {journal} {\bibinfo  {journal} {npj Quantum Information}\
  }\textbf {\bibinfo {volume} {2}},\ \bibinfo {pages} {16002} (\bibinfo {year}
  {2016})},\ \Eprint {http://arxiv.org/abs/1509.01127} {arXiv:1509.01127}
  \BibitemShut {NoStop}%
\bibitem [{\citenamefont {Rosenberg}\ \emph {et~al.}(2017)\citenamefont
  {Rosenberg}, \citenamefont {Kim}, \citenamefont {Das}, \citenamefont {Yost},
  \citenamefont {Gustavsson}, \citenamefont {Hover}, \citenamefont {Krantz},
  \citenamefont {Melville}, \citenamefont {Racz}, \citenamefont {Samach},
  \citenamefont {Weber}, \citenamefont {Yan}, \citenamefont {Yoder},
  \citenamefont {Kerman},\ and\ \citenamefont {Oliver}}]{Rosenberg2017}%
  \BibitemOpen
  \bibfield  {author} {\bibinfo {author} {\bibfnamefont {D.}~\bibnamefont
  {Rosenberg}}, \bibinfo {author} {\bibfnamefont {D.}~\bibnamefont {Kim}},
  \bibinfo {author} {\bibfnamefont {R.}~\bibnamefont {Das}}, \bibinfo {author}
  {\bibfnamefont {D.}~\bibnamefont {Yost}}, \bibinfo {author} {\bibfnamefont
  {S.}~\bibnamefont {Gustavsson}}, \bibinfo {author} {\bibfnamefont
  {D.}~\bibnamefont {Hover}}, \bibinfo {author} {\bibfnamefont
  {P.}~\bibnamefont {Krantz}}, \bibinfo {author} {\bibfnamefont
  {A.}~\bibnamefont {Melville}}, \bibinfo {author} {\bibfnamefont
  {L.}~\bibnamefont {Racz}}, \bibinfo {author} {\bibfnamefont {G.~O.}\
  \bibnamefont {Samach}}, \bibinfo {author} {\bibfnamefont {S.~J.}\
  \bibnamefont {Weber}}, \bibinfo {author} {\bibfnamefont {F.}~\bibnamefont
  {Yan}}, \bibinfo {author} {\bibfnamefont {J.~L.}\ \bibnamefont {Yoder}},
  \bibinfo {author} {\bibfnamefont {A.~J.}\ \bibnamefont {Kerman}}, \ and\
  \bibinfo {author} {\bibfnamefont {W.~D.}\ \bibnamefont {Oliver}},\ }\href
  {\doibase 10.1038/s41534-017-0044-0} {\bibfield  {journal} {\bibinfo
  {journal} {npj Quantum Information}\ }\textbf {\bibinfo {volume} {3}},\
  \bibinfo {pages} {42} (\bibinfo {year} {2017})},\ \Eprint
  {http://arxiv.org/abs/1706.04116} {arXiv:1706.04116} \BibitemShut {NoStop}%
\bibitem [{\citenamefont {Versluis}\ \emph {et~al.}(2017)\citenamefont
  {Versluis}, \citenamefont {Poletto}, \citenamefont {Khammassi}, \citenamefont
  {Tarasinski}, \citenamefont {Haider}, \citenamefont {Michalak}, \citenamefont
  {Bruno}, \citenamefont {Bertels},\ and\ \citenamefont
  {DiCarlo}}]{Versluis2017}%
  \BibitemOpen
  \bibfield  {author} {\bibinfo {author} {\bibfnamefont {R.}~\bibnamefont
  {Versluis}}, \bibinfo {author} {\bibfnamefont {S.}~\bibnamefont {Poletto}},
  \bibinfo {author} {\bibfnamefont {N.}~\bibnamefont {Khammassi}}, \bibinfo
  {author} {\bibfnamefont {B.}~\bibnamefont {Tarasinski}}, \bibinfo {author}
  {\bibfnamefont {N.}~\bibnamefont {Haider}}, \bibinfo {author} {\bibfnamefont
  {D.~J.}\ \bibnamefont {Michalak}}, \bibinfo {author} {\bibfnamefont
  {A.}~\bibnamefont {Bruno}}, \bibinfo {author} {\bibfnamefont
  {K.}~\bibnamefont {Bertels}}, \ and\ \bibinfo {author} {\bibfnamefont
  {L.}~\bibnamefont {DiCarlo}},\ }\href {\doibase
  10.1103/PhysRevApplied.8.034021} {\bibfield  {journal} {\bibinfo  {journal}
  {Physical Review Applied}\ }\textbf {\bibinfo {volume} {8}},\ \bibinfo
  {pages} {034021} (\bibinfo {year} {2017})}\BibitemShut {NoStop}%
\bibitem [{\citenamefont {Naik}\ \emph {et~al.}(2017)\citenamefont {Naik},
  \citenamefont {Leung}, \citenamefont {Chakram}, \citenamefont {Groszkowski},
  \citenamefont {Lu}, \citenamefont {Earnest}, \citenamefont {McKay},
  \citenamefont {Koch},\ and\ \citenamefont {Schuster}}]{Naik2017}%
  \BibitemOpen
  \bibfield  {author} {\bibinfo {author} {\bibfnamefont {R.~K.}\ \bibnamefont
  {Naik}}, \bibinfo {author} {\bibfnamefont {N.}~\bibnamefont {Leung}},
  \bibinfo {author} {\bibfnamefont {S.}~\bibnamefont {Chakram}}, \bibinfo
  {author} {\bibfnamefont {P.}~\bibnamefont {Groszkowski}}, \bibinfo {author}
  {\bibfnamefont {Y.}~\bibnamefont {Lu}}, \bibinfo {author} {\bibfnamefont
  {N.}~\bibnamefont {Earnest}}, \bibinfo {author} {\bibfnamefont {D.~C.}\
  \bibnamefont {McKay}}, \bibinfo {author} {\bibfnamefont {J.}~\bibnamefont
  {Koch}}, \ and\ \bibinfo {author} {\bibfnamefont {D.~I.}\ \bibnamefont
  {Schuster}},\ }\href {\doibase 10.1038/s41467-017-02046-6} {\bibfield
  {journal} {\bibinfo  {journal} {Nature Communications}\ }\textbf {\bibinfo
  {volume} {8}},\ \bibinfo {pages} {1904} (\bibinfo {year} {2017})}\BibitemShut
  {NoStop}%
\bibitem [{\citenamefont {{Charles James
  Neill}}(2017)}]{CharlesJamesNeill2017}%
  \BibitemOpen
  \bibfield  {author} {\bibinfo {author} {\bibnamefont {{Charles James
  Neill}}},\ }\href@noop {} {\emph {\bibinfo {title} {Thesis}}},\ \bibinfo
  {type} {Tech. Rep.}\ (\bibinfo {year} {2017})\BibitemShut {NoStop}%
\bibitem [{\citenamefont {Yan}\ \emph {et~al.}(2018)\citenamefont {Yan},
  \citenamefont {Krantz}, \citenamefont {Sung}, \citenamefont {Kjaergaard},
  \citenamefont {Campbell}, \citenamefont {Orlando}, \citenamefont
  {Gustavsson},\ and\ \citenamefont {Oliver}}]{Yan2018}%
  \BibitemOpen
  \bibfield  {author} {\bibinfo {author} {\bibfnamefont {F.}~\bibnamefont
  {Yan}}, \bibinfo {author} {\bibfnamefont {P.}~\bibnamefont {Krantz}},
  \bibinfo {author} {\bibfnamefont {Y.}~\bibnamefont {Sung}}, \bibinfo {author}
  {\bibfnamefont {M.}~\bibnamefont {Kjaergaard}}, \bibinfo {author}
  {\bibfnamefont {D.~L.}\ \bibnamefont {Campbell}}, \bibinfo {author}
  {\bibfnamefont {T.~P.}\ \bibnamefont {Orlando}}, \bibinfo {author}
  {\bibfnamefont {S.}~\bibnamefont {Gustavsson}}, \ and\ \bibinfo {author}
  {\bibfnamefont {W.~D.}\ \bibnamefont {Oliver}},\ }\href {\doibase
  10.1103/PhysRevApplied.10.054062} {\bibfield  {journal} {\bibinfo  {journal}
  {Physical Review Applied}\ }\textbf {\bibinfo {volume} {10}},\ \bibinfo
  {pages} {054062} (\bibinfo {year} {2018})},\ \Eprint
  {http://arxiv.org/abs/1803.09813} {arXiv:1803.09813} \BibitemShut {NoStop}%
\bibitem [{\citenamefont {Minev}\ \emph {et~al.}(2019)\citenamefont {Minev},
  \citenamefont {Mundhada}, \citenamefont {Shankar}, \citenamefont {Reinhold},
  \citenamefont {Guti{\'{e}}rrez-J{\'{a}}uregui}, \citenamefont {Schoelkopf},
  \citenamefont {Mirrahimi}, \citenamefont {Carmichael},\ and\ \citenamefont
  {Devoret}}]{Minev2019Nature}%
  \BibitemOpen
  \bibfield  {author} {\bibinfo {author} {\bibfnamefont {Z.~K.}\ \bibnamefont
  {Minev}}, \bibinfo {author} {\bibfnamefont {S.~O.}\ \bibnamefont {Mundhada}},
  \bibinfo {author} {\bibfnamefont {S.}~\bibnamefont {Shankar}}, \bibinfo
  {author} {\bibfnamefont {P.}~\bibnamefont {Reinhold}}, \bibinfo {author}
  {\bibfnamefont {R.}~\bibnamefont {Guti{\'{e}}rrez-J{\'{a}}uregui}}, \bibinfo
  {author} {\bibfnamefont {R.~J.}\ \bibnamefont {Schoelkopf}}, \bibinfo
  {author} {\bibfnamefont {M.}~\bibnamefont {Mirrahimi}}, \bibinfo {author}
  {\bibfnamefont {H.~J.}\ \bibnamefont {Carmichael}}, \ and\ \bibinfo {author}
  {\bibfnamefont {M.~H.}\ \bibnamefont {Devoret}},\ }\href {\doibase
  10.1038/s41586-019-1287-z} {\bibfield  {journal} {\bibinfo  {journal}
  {Nature}\ }\textbf {\bibinfo {volume} {570}},\ \bibinfo {pages} {200}
  (\bibinfo {year} {2019})},\ \Eprint {http://arxiv.org/abs/1803.00545}
  {arXiv:1803.00545} \BibitemShut {NoStop}%
\bibitem [{\citenamefont {{Antonio D.}}\ \emph {et~al.}(2021)\citenamefont
  {{Antonio D.}}, \citenamefont {Takita}, \citenamefont {Inoue}, \citenamefont
  {Lekuch}, \citenamefont {Minev}, \citenamefont {Chow},\ and\ \citenamefont
  {Gambetta}}]{Corcoles2021}%
  \BibitemOpen
  \bibfield  {author} {\bibinfo {author} {\bibnamefont {{Antonio D.}}},
  \bibinfo {author} {\bibfnamefont {M.}~\bibnamefont {Takita}}, \bibinfo
  {author} {\bibfnamefont {K.}~\bibnamefont {Inoue}}, \bibinfo {author}
  {\bibfnamefont {S.}~\bibnamefont {Lekuch}}, \bibinfo {author} {\bibfnamefont
  {Z.~K.}\ \bibnamefont {Minev}}, \bibinfo {author} {\bibfnamefont {J.~M.}\
  \bibnamefont {Chow}}, \ and\ \bibinfo {author} {\bibfnamefont {J.~M.}\
  \bibnamefont {Gambetta}},\ }\href {http://arxiv.org/abs/2102.01682} {\
  (\bibinfo {year} {2021})},\ \Eprint {http://arxiv.org/abs/2102.01682}
  {arXiv:2102.01682} \BibitemShut {NoStop}%
\bibitem [{\citenamefont {Yurke}\ and\ \citenamefont
  {Denker}(1984)}]{Yurke1984}%
  \BibitemOpen
  \bibfield  {author} {\bibinfo {author} {\bibfnamefont {B.}~\bibnamefont
  {Yurke}}\ and\ \bibinfo {author} {\bibfnamefont {J.~S.}\ \bibnamefont
  {Denker}},\ }\href {\doibase 10.1103/PhysRevA.29.1419} {\bibfield  {journal}
  {\bibinfo  {journal} {Physical Review A}\ }\textbf {\bibinfo {volume} {29}},\
  \bibinfo {pages} {1419} (\bibinfo {year} {1984})}\BibitemShut {NoStop}%
\bibitem [{\citenamefont {Devoret}(1995)}]{Devoret1995}%
  \BibitemOpen
  \bibfield  {author} {\bibinfo {author} {\bibfnamefont {M.~H.}\ \bibnamefont
  {Devoret}},\ }in\ \href@noop {} {\emph {\bibinfo {booktitle} {Quantum
  Fluctuations, Les Houches, Sess. LXIII}}}\ (\bibinfo {year}
  {1995})\BibitemShut {NoStop}%
\bibitem [{\citenamefont {Burkard}\ \emph {et~al.}(2004)\citenamefont
  {Burkard}, \citenamefont {Koch},\ and\ \citenamefont
  {DiVincenzo}}]{Burkard2004}%
  \BibitemOpen
  \bibfield  {author} {\bibinfo {author} {\bibfnamefont {G.}~\bibnamefont
  {Burkard}}, \bibinfo {author} {\bibfnamefont {R.~H.}\ \bibnamefont {Koch}}, \
  and\ \bibinfo {author} {\bibfnamefont {D.~P.}\ \bibnamefont {DiVincenzo}},\
  }\href {\doibase 10.1103/PhysRevB.69.064503} {\bibfield  {journal} {\bibinfo
  {journal} {Physical Review B}\ }\textbf {\bibinfo {volume} {69}},\ \bibinfo
  {pages} {064503} (\bibinfo {year} {2004})}\BibitemShut {NoStop}%
\bibitem [{\citenamefont {Koch}\ \emph {et~al.}(2007)\citenamefont {Koch},
  \citenamefont {Yu}, \citenamefont {Gambetta}, \citenamefont {Houck},
  \citenamefont {Schuster}, \citenamefont {Majer}, \citenamefont {Blais},
  \citenamefont {Devoret}, \citenamefont {Girvin},\ and\ \citenamefont
  {Schoelkopf}}]{Koch2007}%
  \BibitemOpen
  \bibfield  {author} {\bibinfo {author} {\bibfnamefont {J.}~\bibnamefont
  {Koch}}, \bibinfo {author} {\bibfnamefont {T.~M.}\ \bibnamefont {Yu}},
  \bibinfo {author} {\bibfnamefont {J.}~\bibnamefont {Gambetta}}, \bibinfo
  {author} {\bibfnamefont {A.~A.}\ \bibnamefont {Houck}}, \bibinfo {author}
  {\bibfnamefont {D.~I.}\ \bibnamefont {Schuster}}, \bibinfo {author}
  {\bibfnamefont {J.}~\bibnamefont {Majer}}, \bibinfo {author} {\bibfnamefont
  {A.}~\bibnamefont {Blais}}, \bibinfo {author} {\bibfnamefont {M.~H.}\
  \bibnamefont {Devoret}}, \bibinfo {author} {\bibfnamefont {S.~M.}\
  \bibnamefont {Girvin}}, \ and\ \bibinfo {author} {\bibfnamefont {R.~J.}\
  \bibnamefont {Schoelkopf}},\ }\href {\doibase 10.1103/PhysRevA.76.042319}
  {\bibfield  {journal} {\bibinfo  {journal} {Physical Review A}\ }\textbf
  {\bibinfo {volume} {76}},\ \bibinfo {pages} {42319} (\bibinfo {year}
  {2007})}\BibitemShut {NoStop}%
\bibitem [{Note1()}]{Note1}%
  \BibitemOpen
  \bibinfo {note} {For an early version of the open-source code \protect \citep
  {Qiskit_Metal} developed by the authors of this manuscript, see \protect
  \href {http://www.qiskit.org/metal}{qiskit.org/metal}. The Qiskit Metal
  project builds on \protect \href
  {http://http:/github.com/zlatko-minev/pyEPR}{github.com/zlatko-minev/pyEPR}.}\BibitemShut
  {Stop}%
\bibitem [{\citenamefont {Zimmerman}\ and\ \citenamefont
  {Silver}(1966)}]{Zimmerman1966}%
  \BibitemOpen
  \bibfield  {author} {\bibinfo {author} {\bibfnamefont {J.~E.}\ \bibnamefont
  {Zimmerman}}\ and\ \bibinfo {author} {\bibfnamefont {A.~H.}\ \bibnamefont
  {Silver}},\ }\href {\doibase 10.1103/PhysRev.141.367} {\bibfield  {journal}
  {\bibinfo  {journal} {Physical Review}\ }\textbf {\bibinfo {volume} {141}},\
  \bibinfo {pages} {367} (\bibinfo {year} {1966})}\BibitemShut {NoStop}%
\bibitem [{\citenamefont {Clarke}\ and\ \citenamefont
  {Braginski}(2004)}]{Clarke2004}%
  \BibitemOpen
  \bibinfo {editor} {\bibfnamefont {J.}~\bibnamefont {Clarke}}\ and\ \bibinfo
  {editor} {\bibfnamefont {A.~I.}\ \bibnamefont {Braginski}},\ eds.,\ \href
  {\doibase 10.1002/3527603646} {\emph {\bibinfo {title} {{The SQUID
  Handbook}}}}\ (\bibinfo  {publisher} {Wiley-VCH Verlag GmbH {\&} Co. KGaA},\
  \bibinfo {address} {Weinheim, FRG},\ \bibinfo {year} {2004})\BibitemShut
  {NoStop}%
\bibitem [{\citenamefont {Frattini}\ \emph {et~al.}(2018)\citenamefont
  {Frattini}, \citenamefont {Sivak}, \citenamefont {Lingenfelter},
  \citenamefont {Shankar},\ and\ \citenamefont {Devoret}}]{Frattini2018}%
  \BibitemOpen
  \bibfield  {author} {\bibinfo {author} {\bibfnamefont {N.~E.}\ \bibnamefont
  {Frattini}}, \bibinfo {author} {\bibfnamefont {V.~V.}\ \bibnamefont {Sivak}},
  \bibinfo {author} {\bibfnamefont {A.}~\bibnamefont {Lingenfelter}}, \bibinfo
  {author} {\bibfnamefont {S.}~\bibnamefont {Shankar}}, \ and\ \bibinfo
  {author} {\bibfnamefont {M.~H.}\ \bibnamefont {Devoret}},\ }\href {\doibase
  10.1103/PhysRevApplied.10.054020} {\bibfield  {journal} {\bibinfo  {journal}
  {Physical Review Applied}\ } (\bibinfo {year} {2018}),\
  10.1103/PhysRevApplied.10.054020},\ \Eprint {http://arxiv.org/abs/1806.06093}
  {arXiv:1806.06093} \BibitemShut {NoStop}%
\bibitem [{\citenamefont {Minev}(2019)}]{Minev2019-Thesis}%
  \BibitemOpen
  \bibfield  {author} {\bibinfo {author} {\bibfnamefont {Z.~K.}\ \bibnamefont
  {Minev}},\ }\href {http://arxiv.org/abs/1902.10355} {\bibfield  {journal}
  {\bibinfo  {journal} {Ph.D. thesis, Yale University}\ } (\bibinfo {year}
  {2019})},\ \Eprint {http://arxiv.org/abs/1902.10355} {arXiv:1902.10355}
  \BibitemShut {NoStop}%
\bibitem [{\citenamefont {Gloos}\ \emph {et~al.}(2000)\citenamefont {Gloos},
  \citenamefont {Poikolainen},\ and\ \citenamefont {Pekola}}]{Gloos2000}%
  \BibitemOpen
  \bibfield  {author} {\bibinfo {author} {\bibfnamefont {K.}~\bibnamefont
  {Gloos}}, \bibinfo {author} {\bibfnamefont {R.~S.}\ \bibnamefont
  {Poikolainen}}, \ and\ \bibinfo {author} {\bibfnamefont {J.~P.}\ \bibnamefont
  {Pekola}},\ }\href {\doibase 10.1063/1.1320861} {\bibfield  {journal}
  {\bibinfo  {journal} {Applied Physics Letters}\ }\textbf {\bibinfo {volume}
  {77}},\ \bibinfo {pages} {2915} (\bibinfo {year} {2000})}\BibitemShut
  {NoStop}%
\bibitem [{\citenamefont {Wolff}(2006)}]{Wolff2006}%
  \BibitemOpen
  \bibfield  {author} {\bibinfo {author} {\bibfnamefont {I.}~\bibnamefont
  {Wolff}},\ }\href {\doibase 10.1002/0470040882} {\emph {\bibinfo {title}
  {Coplanar Microwave Integrated Circuits}}}\ (\bibinfo  {publisher} {John
  Wiley {\&} Sons, Inc.},\ \bibinfo {address} {Hoboken, NJ, USA},\ \bibinfo
  {year} {2006})\BibitemShut {NoStop}%
\bibitem [{\citenamefont {Dirac}(1982)}]{Dirac1982-Book}%
  \BibitemOpen
  \bibfield  {author} {\bibinfo {author} {\bibfnamefont {P.~A.~M.}\
  \bibnamefont {Dirac}},\ }\href
  {https://www.valorebooks.com/textbooks/principles-of-quantum-mechanics/9780198520115?gclid=Cj0KCQiA2o{\_}fBRC8ARIsAIOyQ-nfcm-XiKsjvyzk4TapaXokV6wpqrl7gRKZOi2sMYyNs6apjtU9X7AaAvLWEALw{\_}wcB{\&}gclsrc=aw.ds}
  {\emph {\bibinfo {title} {{Principles of Quantum Mechanics}}}}\ (\bibinfo
  {publisher} {Oxford University Press},\ \bibinfo {year} {1982})\BibitemShut
  {NoStop}%
\bibitem [{Note2()}]{Note2}%
  \BibitemOpen
  \bibinfo {note} {This is the form employed in a recent package entitled
  \protect \href
  {https://scqubits.readthedocs.io/en/latest/index.html}{scqubits}; see also
  Ref.~\protect \rev@citealpnum {Kerman2020}.}\BibitemShut {Stop}%
\bibitem [{\citenamefont {Minev}\ \emph {et~al.}(2021)\citenamefont {Minev},
  \citenamefont {McConkey}, \citenamefont {Drysdal}, \citenamefont {Shah},
  \citenamefont {Wang}, \citenamefont {Facchini}, \citenamefont {Harper},
  \citenamefont {Blair}, \citenamefont {Zhang}, \citenamefont {Lanzillo},
  \citenamefont {Mukesh}, \citenamefont {Shanks}, \citenamefont {Warren},\ and\
  \citenamefont {Gambetta}}]{Qiskit_Metal}%
  \BibitemOpen
  \bibfield  {author} {\bibinfo {author} {\bibfnamefont {Z.~K.}\ \bibnamefont
  {Minev}}, \bibinfo {author} {\bibfnamefont {T.~G.}\ \bibnamefont {McConkey}},
  \bibinfo {author} {\bibfnamefont {J.}~\bibnamefont {Drysdal}}, \bibinfo
  {author} {\bibfnamefont {P.}~\bibnamefont {Shah}}, \bibinfo {author}
  {\bibfnamefont {D.}~\bibnamefont {Wang}}, \bibinfo {author} {\bibfnamefont
  {M.}~\bibnamefont {Facchini}}, \bibinfo {author} {\bibfnamefont
  {G.}~\bibnamefont {Harper}}, \bibinfo {author} {\bibfnamefont
  {J.}~\bibnamefont {Blair}}, \bibinfo {author} {\bibfnamefont
  {H.}~\bibnamefont {Zhang}}, \bibinfo {author} {\bibfnamefont
  {N.}~\bibnamefont {Lanzillo}}, \bibinfo {author} {\bibfnamefont
  {S.}~\bibnamefont {Mukesh}}, \bibinfo {author} {\bibfnamefont
  {W.}~\bibnamefont {Shanks}}, \bibinfo {author} {\bibfnamefont
  {C.}~\bibnamefont {Warren}}, \ and\ \bibinfo {author} {\bibfnamefont {J.~M.}\
  \bibnamefont {Gambetta}},\ }\href@noop {} {\enquote {\bibinfo {title}
  {{Qiskit Metal: An Open-Source Framework for Quantum Device Design {\&}
  Analysis}},}\ } (\bibinfo {year} {2021})\BibitemShut {NoStop}%
\bibitem [{\citenamefont {Zangwill}(2012)}]{zangwill2012-book}%
  \BibitemOpen
  \bibfield  {author} {\bibinfo {author} {\bibfnamefont {A.}~\bibnamefont
  {Zangwill}},\ }\href {\doibase 10.1017/CBO9781139034777} {\emph {\bibinfo
  {title} {{Modern Electrodynamics}}}}\ (\bibinfo  {publisher} {Cambridge
  University Press},\ \bibinfo {year} {2012})\BibitemShut {NoStop}%
\bibitem [{\citenamefont {Blais}\ \emph {et~al.}(2004)\citenamefont {Blais},
  \citenamefont {Huang}, \citenamefont {Wallraff}, \citenamefont {Girvin},\
  and\ \citenamefont {Schoelkopf}}]{Blais2004}%
  \BibitemOpen
  \bibfield  {author} {\bibinfo {author} {\bibfnamefont {A.}~\bibnamefont
  {Blais}}, \bibinfo {author} {\bibfnamefont {R.-S.}\ \bibnamefont {Huang}},
  \bibinfo {author} {\bibfnamefont {A.}~\bibnamefont {Wallraff}}, \bibinfo
  {author} {\bibfnamefont {S.~M.}\ \bibnamefont {Girvin}}, \ and\ \bibinfo
  {author} {\bibfnamefont {R.~J.}\ \bibnamefont {Schoelkopf}},\ }\href
  {\doibase 10.1103/PhysRevA.69.062320} {\bibfield  {journal} {\bibinfo
  {journal} {Physical Review A}\ }\textbf {\bibinfo {volume} {69}},\ \bibinfo
  {pages} {062320} (\bibinfo {year} {2004})},\ \Eprint
  {http://arxiv.org/abs/0402216} {arXiv:0402216 [cond-mat]} \BibitemShut
  {NoStop}%
\bibitem [{\citenamefont {Gambetta}\ \emph {et~al.}(2006)\citenamefont
  {Gambetta}, \citenamefont {Blais}, \citenamefont {Schuster}, \citenamefont
  {Wallraff}, \citenamefont {Frunzio}, \citenamefont {Majer}, \citenamefont
  {Devoret}, \citenamefont {Girvin},\ and\ \citenamefont
  {Schoelkopf}}]{Gambetta2006}%
  \BibitemOpen
  \bibfield  {author} {\bibinfo {author} {\bibfnamefont {J.}~\bibnamefont
  {Gambetta}}, \bibinfo {author} {\bibfnamefont {A.}~\bibnamefont {Blais}},
  \bibinfo {author} {\bibfnamefont {D.~I.}\ \bibnamefont {Schuster}}, \bibinfo
  {author} {\bibfnamefont {A.}~\bibnamefont {Wallraff}}, \bibinfo {author}
  {\bibfnamefont {L.}~\bibnamefont {Frunzio}}, \bibinfo {author} {\bibfnamefont
  {J.}~\bibnamefont {Majer}}, \bibinfo {author} {\bibfnamefont {M.~H.}\
  \bibnamefont {Devoret}}, \bibinfo {author} {\bibfnamefont {S.~M.}\
  \bibnamefont {Girvin}}, \ and\ \bibinfo {author} {\bibfnamefont {R.~J.}\
  \bibnamefont {Schoelkopf}},\ }\href {\doibase 10.1103/PhysRevA.74.042318}
  {\bibfield  {journal} {\bibinfo  {journal} {Physical Review A - Atomic,
  Molecular, and Optical Physics}\ } (\bibinfo {year} {2006}),\
  10.1103/PhysRevA.74.042318},\ \Eprint {http://arxiv.org/abs/0602322}
  {arXiv:0602322 [cond-mat]} \BibitemShut {NoStop}%
\bibitem [{\citenamefont {Gambetta}\ \emph {et~al.}(2008)\citenamefont
  {Gambetta}, \citenamefont {Blais}, \citenamefont {Boissonneault},
  \citenamefont {Houck}, \citenamefont {Schuster},\ and\ \citenamefont
  {Girvin}}]{Gambetta2008-qm-traj}%
  \BibitemOpen
  \bibfield  {author} {\bibinfo {author} {\bibfnamefont {J.}~\bibnamefont
  {Gambetta}}, \bibinfo {author} {\bibfnamefont {A.}~\bibnamefont {Blais}},
  \bibinfo {author} {\bibfnamefont {M.}~\bibnamefont {Boissonneault}}, \bibinfo
  {author} {\bibfnamefont {A.~A.}\ \bibnamefont {Houck}}, \bibinfo {author}
  {\bibfnamefont {D.~I.}\ \bibnamefont {Schuster}}, \ and\ \bibinfo {author}
  {\bibfnamefont {S.~M.}\ \bibnamefont {Girvin}},\ }\href {\doibase
  10.1103/PhysRevA.77.012112} {\bibfield  {journal} {\bibinfo  {journal}
  {Physical Review A}\ }\textbf {\bibinfo {volume} {77}},\ \bibinfo {pages}
  {012112} (\bibinfo {year} {2008})}\BibitemShut {NoStop}%
\bibitem [{\citenamefont {Krupka}\ \emph {et~al.}(2006)\citenamefont {Krupka},
  \citenamefont {Breeze}, \citenamefont {Centeno}, \citenamefont {Alford},
  \citenamefont {Claussen},\ and\ \citenamefont {Jensen}}]{Krupka2006}%
  \BibitemOpen
  \bibfield  {author} {\bibinfo {author} {\bibfnamefont {J.}~\bibnamefont
  {Krupka}}, \bibinfo {author} {\bibfnamefont {J.}~\bibnamefont {Breeze}},
  \bibinfo {author} {\bibfnamefont {A.}~\bibnamefont {Centeno}}, \bibinfo
  {author} {\bibfnamefont {N.}~\bibnamefont {Alford}}, \bibinfo {author}
  {\bibfnamefont {T.}~\bibnamefont {Claussen}}, \ and\ \bibinfo {author}
  {\bibfnamefont {L.}~\bibnamefont {Jensen}},\ }\href {\doibase
  10.1109/TMTT.2006.883655} {\bibfield  {journal} {\bibinfo  {journal} {IEEE
  Transactions on Microwave Theory and Techniques}\ }\textbf {\bibinfo {volume}
  {54}},\ \bibinfo {pages} {3995} (\bibinfo {year} {2006})}\BibitemShut
  {NoStop}%
\bibitem [{\citenamefont {Simons}(2001)}]{Simons2001}%
  \BibitemOpen
  \bibfield  {author} {\bibinfo {author} {\bibfnamefont {R.~N.}\ \bibnamefont
  {Simons}},\ }\href {\doibase 10.1002/0471224758} {\emph {\bibinfo {title}
  {Coplanar Waveguide Circuits, Components, and Systems}}}\ (\bibinfo {year}
  {2001})\BibitemShut {NoStop}%
\bibitem [{\citenamefont {Minev}\ \emph {et~al.}(2016)\citenamefont {Minev},
  \citenamefont {Serniak}, \citenamefont {Pop}, \citenamefont {Leghtas},
  \citenamefont {Sliwa}, \citenamefont {Hatridge}, \citenamefont {Frunzio},
  \citenamefont {Schoelkopf},\ and\ \citenamefont {Devoret}}]{Minev2016}%
  \BibitemOpen
  \bibfield  {author} {\bibinfo {author} {\bibfnamefont {Z.}~\bibnamefont
  {Minev}}, \bibinfo {author} {\bibfnamefont {K.}~\bibnamefont {Serniak}},
  \bibinfo {author} {\bibfnamefont {I.~M.}\ \bibnamefont {Pop}}, \bibinfo
  {author} {\bibfnamefont {Z.}~\bibnamefont {Leghtas}}, \bibinfo {author}
  {\bibfnamefont {K.}~\bibnamefont {Sliwa}}, \bibinfo {author} {\bibfnamefont
  {M.}~\bibnamefont {Hatridge}}, \bibinfo {author} {\bibfnamefont
  {L.}~\bibnamefont {Frunzio}}, \bibinfo {author} {\bibfnamefont {R.~J.}\
  \bibnamefont {Schoelkopf}}, \ and\ \bibinfo {author} {\bibfnamefont {M.~H.}\
  \bibnamefont {Devoret}},\ }\href {\doibase 10.1103/PhysRevApplied.5.044021}
  {\bibfield  {journal} {\bibinfo  {journal} {Physical Review Applied}\
  }\textbf {\bibinfo {volume} {5}},\ \bibinfo {pages} {044021} (\bibinfo {year}
  {2016})},\ \Eprint {http://arxiv.org/abs/1509.01619} {arXiv:1509.01619}
  \BibitemShut {NoStop}%
\end{thebibliography}%

\end{document}